\documentclass[12pt, a4paper]{article}
 \usepackage[font=small,format=plain,labelfont=bf,up,textfont=normal,up,justification=justified,singlelinecheck=false]{caption}
\usepackage{subcaption}
\usepackage{mathrsfs}
\usepackage[a4paper, left=2cm,right=2cm]{geometry}
\usepackage[colorlinks=true,linkcolor=black,citecolor=teal,urlcolor=MidnightBlue,filecolor=black]{hyperref}
\usepackage{amsfonts}
\usepackage{amsmath,amssymb}
\usepackage{setspace}
\usepackage{slashed}
\usepackage{braket}
\usepackage{color}
\usepackage{amsmath}
\usepackage[dvipsnames]{xcolor}
\definecolor{SchoolColor}{rgb}{0.6471, 0.1098, 0.1882} 
\usepackage{subfloat}

\usepackage[utf8,applemac]{inputenc}
\usepackage{tensor}
\usepackage{cite}
\usepackage{tikz}
\usetikzlibrary{calc}
\usetikzlibrary{patterns}
\usetikzlibrary{arrows.meta}

\usepackage{graphicx}
\usepackage{bm} 

\usepackage[utf8,applemac]{inputenc}
\usepackage{tensor}
\usepackage{cite}
\usepackage{tikz}
\usepackage{graphicx}
\graphicspath{{figure/}}
\usepackage{array}
\usepackage{booktabs}

\usepackage{multirow}

\usepackage{dcolumn}
\usepackage{bm}

\usepackage{verbatim}

\usepackage{textcomp} 
\usepackage{graphicx} 

\numberwithin{equation}{section}
\newcommand{\bea}{\begin{eqnarray}}
\newcommand{\eea}{\end{eqnarray}}
\newcommand{\be}{\begin{equation}}
\newcommand{\ee}{\end{equation}}
\newcommand{\bs}{\begin{subequations}}
\newcommand{\es}{\end{subequations}}
\def\nn{\nonumber}

\newcommand{\beqs}{\begin{eqnarray}}
\newcommand{\eeqs}{\end{eqnarray}}
\numberwithin{equation}{section}

\newcommand{\Rmnum}[1]{\uppercase\expandafter{\romannumeral #1\relax}}

\usepackage{subcaption}
\usepackage{subcaption}

\newcommand{\yrz}{\color{red}}

\begin{document}
\begin{titlepage}

\begin{flushright}\vspace{-3cm}
{\small
\today }\end{flushright}
\vspace{0.5cm}
\begin{center}
	{{ \LARGE{\bf{
Gravitational helicity flux density from two-body systems  }}}}\vspace{5mm}

	\centerline{\large{\bf 
	Jiang Long\footnote{
				longjiang@hust.edu.cn} and Run-Ze Yu\footnote{yurunze01@hust.edu.cn}}}
	\vspace{2mm}
	\normalsize
	\bigskip\medskip

	\textit{School of Physics, Huazhong University of Science and Technology, \\ Luoyu Road 1037, Wuhan, Hubei 430074, China
	}
	
	\vspace{25mm}
	
	\begin{abstract}
		\noindent
		{The helicity flux density is a novel quantity which characterizes the angle-dependence of the helicity of radiative gravitons and it may be tested by gravitational wave experiments in the future. We derive a quadrupole formula for the helicity flux density due to gravitational radiation in the slow motion and the weak field limit. We apply the formula to the bound and  unbound orbits in two-body systems and find that the total radiative helicity fluxes are always zero. However, the radiative helicity flux density, which is $\mathcal{O}(G^3)$ in the Newtonian limit,  still has non-trivial dependence  on the angle. Furthermore, we also find a  formula for the total helicity flux   by including all contributions of the higher multipoles.}\end{abstract}
	

\end{center}

\end{titlepage}
\tableofcontents

\section{Introduction}
Gravitational wave, one of the great predictions of general relativity,  were detected several years ago \cite{LIGOScientific:2016aoc}. It is well known that gravitational waves carry energy, linear momentum as well as angular momentum during their propagation. The energy loss due to gravitational radiation,  which is governed by the famous formula \cite{Bondi:1962px}
\begin{align}
\frac{dE}{dud\Omega}=-T(u,\Omega),\quad T(u,\Omega)=\frac{1}{32\pi G}\dot{C}_{AB}\dot{C}^{AB},\label{bondiloss}
\end{align} 
was observed indirectly in a pulsar binary system, PSR B1913+16 by Hulse and Taylor 50 years ago \cite{Hulse:1974eb,1982ApJ...253..908T}. 
In the above formula, the so-called energy flux density $T\left(u,\Omega\right) $ is defined at the future null infinity $\mathcal{I}^+$ with the coordinate $u$ the retarded time and 
$\Omega$ the spherical coordinates of the celestial sphere.
The shear tensor $C_{AB}$ encodes information about gravitational waves, and its time  derivative $\dot{C}_{AB}=\frac{\partial}{\partial u}C_{AB}$ represents the Bondi news tensor. It is worth noting that the shear tensor $C_{AB}$ is both symmetric and traceless. We adopt the convention that capital Roman indices $A,B,\cdots$ are raised and lowered with the metric $\gamma_{AB}$ of the unit sphere, and the  Greek letter indices $\mu, \nu,\cdots$ are raised and lowered with the metric $\eta_{\mu\nu}$.
\par
Recently, a new 
operator $O(u,\Omega)$, which is called helicity flux density, has been derived in the  context of flat holography \cite{Liu:2023gwa,Liu:2024nkc}
\begin{align} 
O(u,\Omega)=\frac{1}{32\pi G}\dot{C}_{AB}C^B_{\ C}\epsilon^{CA},\label{helicityO}
\end{align} where $\epsilon^{AB}$ is the Levi-Civita tensor on the unit sphere.
This operator, arising naturally from the commutator between  generalized superrotation generators \cite{Campiglia:2014yka} and the requirement of the closure of the Lie algebra, is also defined at future null infinity $\mathcal{I}^+$ and obeys the flux equation 
\begin{align} 
\frac{dH}{dud\Omega}=O(u,\Omega),\label{dH}
\end{align} where $H$ is helicity flux across $\mathcal{I}^+$. 
At the microscopic level, the helicity flux evaluates the difference between the
numbers of gravitons with left and right helicity. Therefore, the operator $O(u,\Omega)$ characterizes the rate of change of helicity flux in unit time and unit solid angle. One can define the associated the smeared operator
\begin{align}
\mathcal{O}_g=\int du d\Omega g(\Omega)O(u,\Omega),\label{Og}
\end{align}
  that generates super-duality transformation and  rotates the gravitational  electric-magnetic duality transformations locally.
The super-duality transformation is also called ``dual supertranslation'' \cite{Godazgar:2018qpq, Godazgar:2018dvh,Godazgar:2020gqd} in the literature and contributes to the spin precession of freely falling gyroscopes \cite{Seraj:2022qyt}, an extension of the spin memory effect \cite{Pasterski:2015tva}.
Note that the helicity flux density operator is also related to the ``dual covariant mass aspect'' up to a linear term \cite{Freidel:2021qpz,Freidel:2021ytz}.  It is well known that a massless spinning particle is characterized by the helicity, which is the projection of the particle spin in the direction of its motion. It is a bit surprising that the helicity flux density has not been emphasized during the progress of radiation. The spin of the particle is always intertwined with the angular momentum which makes its independent physical meaning obscure. However, the helicity flux operator counts the difference in the number of massless particles between left and right helicity which slightly extends the concept of the original helicity. Actually, the helicity flux operator generates the super-duality transformation, which was only realized until recently. This may be one of the crucial reasons why people have not tried to construct the helicity flux operator.  

The formula \eqref{helicityO} is valid near $\mathcal{I}^+$ and we should relate it to the source that generates the radiation. At the linear level, assuming the source is far away from the observer, the energy loss is mainly from the variation of the quadrupole \cite{Einstein:1916,Einstein:1918,landau2013classical,thorne2000gravitation}. In \cite{Peters:1963ux}, the energy loss due to gravitational radiation has been discussed for Kepler orbits.  Furthermore, the results have been extended to hyperbolic orbits in \cite{1972PhRvD...5.1021H,1977ApJ...216..610T,Capozziello:2008ra}.  Additionally, the radiation of the linear momentum and the angular momentum have been discussed in \cite{1964PhRv..136.1224P,Thorne:1980ru}. See also recent developments in \cite{Compere:2019gft,Blanchet:2020ngx,Blanchet:2023pce,Siddhant:2024nft}.  Therefore, it is natural to ask for a similar
quadrupole formula for the helicity flux density in gravitational radiation. 

 In this paper, we will  derive the quadrupole formula for the helicity flux density in the weak field and slow motion limit. Though the total helicity flux on the celestial sphere vanishes, the helicity flux density itself is angle-dependent. For periodic  planar orbits, it is shown that the helicity flux density reaches its maximum/minimum at the north/south pole and vanishes on the equatorial plane. The formula can also be applied to the gravitational wave event GW150914 \cite{LIGOScientific:2016aoc}. 

The quadrupole formula (and its higher multipole extension) opens a new window to study gravitational wave astronomy and cosmology. We emphasize that the helicity flux is as important as energy and angular momentum fluxes. The helicity flux density may provide new constraints on the parameters that characterize compact binaries and study the extreme matter such as neutron stars. We may also use the helicity flux density to test general relativity since an alternative gravitational theory may lead to a different helicity flux density. In cosmology,  the Hubble tension \cite{Freedman:2017yms,Feeney:2017sgx,Verde:2019ivm}, a discrepancy between the measurements from CMB and cosmic distance ladder, has attracted a lot of attention. The helicity flux density provides an independent way to define cosmological distance that differs from luminosity distance and thus may contribute to the resolution of  the Hubble tension in the future. 

This paper is organized as follows. In section  \ref{quadrupoleformula} we will derive the quadrupole formula for the helicity flux density. In section \ref{binarysystem}, we will apply the quadrupole formula to various orbits in the two-body systems of astrophysics. Then we will extend the quadrupole formula by including higher multipoles in the following section. A detailed discussion on the application to the gravitational wave event GW150914 is presented in section \ref{application}. We will also introduce a new cosmological distance in the same section. We will discuss our results in section \ref{dis}. The technical details on the integrals on the unit sphere and the quadrupole formula for the planar system are collected in two  appendices.
\section{The quadrupole formula}\label{quadrupoleformula}
In the weak field limit, the metric $g_{\mu\nu}$ around the Minkowski spacetime can be expanded as
\begin{align}
g_{\mu\nu}=\eta_{\mu\nu}+h_{\mu\nu},
\end{align} where $h_{\mu\nu}$ is the perturbation of the gravitational wave and $\eta_{\mu\nu}=\text{diag}(-1,1,1,1)$ is the  Minkowski matrix. We may define a trace-reversed tensor
\begin{align}
    \bar{h}_{\mu\nu}=h_{\mu\nu}-\frac{1}{2}\eta_{\mu\nu}h
\end{align} with $h=\eta^{\mu\nu}h_{\mu\nu}$  the trace of $h_{\mu\nu}$. 
At the linear level, by imposing the de Donder (or harmonic) gauge $\partial^\mu \bar{h}_{\mu\nu}=0$, the trace-reversed perturbation can be solved by Green's function and is determined by the stress tensor $T_{\mu\nu}$ of the source \cite{carroll2019spacetime} 
\begin{align}
    \bar{h}_{\mu\nu}=4G\int d^3\bm x'\frac{T_{\mu\nu}(t-|\bm x-\bm x'|,\bm x')}{|\bm x-\bm x'|}.
\end{align} We assume the source moves slowly and the size $a$ of the source is much smaller than the distance $r=|\bm x|$
\begin{align} a\ll r. 
\end{align} As a consequence, the trace-reversed perturbation is related to the second time derivative of the quadrupole moment 
\begin{align} 
\bar{h}_{ij}=\frac{2G}{r}\ddot{I}_{ij}(u),
\end{align} where $I_{ij}$ is the quadrupole momentum tensor
\begin{align} 
I_{ij}(u)=\int d^3\bm x T_{00}(u,\bm x)x^i x^j.\label{quadrupole}
\end{align} 
The symmetric traceless perturbation $h^{\text{T}}_{ij}=h_{ij}-\frac{1}{3}\delta_{ij}h$ is determined by 
\begin{align}
    h^{\text{T}}_{ij}=\frac{2G}{r}\ddot{M}_{ij}(u),
\end{align} where the reduced quadrupole momentum $M_{ij}$ is symmetric and traceless 
\begin{align}
M_{ij}=I_{ij}-\frac{1}{3}\delta_{ij}I,\quad I=\delta^{kl}I_{kl}.
\end{align}
We may project the symmetric traceless perturbation to the symmetric traceless and transverse mode
\begin{align}
    h^{\text{TT}}_{ij}=\left(P_{i}^k P_j^l-\frac{1}{2}P_{ij}P^{kl}\right)h^{\text{T}}_{kl}\equiv \frac{H^{\text{TT}}_{ij}}{r},
\end{align} where 
\begin{align}
P_{ij}=\delta_{ij}-n_i n_j
\end{align} is the projector  and $n_i$ is the unit normal vector on the sphere 
\begin{align} 
n_i=(\sin\theta\cos\phi,\sin\theta\sin\phi,\cos\theta).
\end{align} 
We may transform the harmonic gauge to the Bondi gauge, the shear tensor is given by  \cite{Blanchet:2020ngx} 
\begin{align}\label{Bondishear}
C_{AB}&=Y^i_AY^j_B H_{ij}^{\text{TT}}=2G(Y^i_AY^j_B+\frac{1}{2}\gamma_{AB}n^in^j)\ddot{M}_{ij}.
\end{align} The vectors $Y^i_A,\ i=1,2,3$ are three conformal Killing vectors which are related to the normal vector by
\begin{align} 
Y^i_A=-\nabla_A n^i.
\end{align} Their explicit expressions and various relevant identities can be found in \cite{Liu:2022mne,Li:2023xrr}.
After some efforts, we find the following quadrupole formula for the helicity flux density
\begin{align}
    \frac{dH}{dud\Omega}=\frac{G}{8\pi}\dddot{M}_{ij}\ddot{M}_{kl}Q^{ijkl},\label{local}
\end{align} where 
\begin{align}
   Q^{ijkl}=&-\delta^{jk}\epsilon^{ilm}n_m+\epsilon^{ilm}n^jn^kn_m-\frac{1}{2}\epsilon^{klm}n^in^jn_m-\frac{1}{2}\epsilon^{ijm}n^kn^ln_m.
\end{align} Note that the last two terms contribute zero since the Levi-Civita tensor is antisymmetric while the reduced quadrupole is symmetric. Therefore, we may drop them and rewrite $Q^{ijkl}$ as 
\be 
Q^{ijkl}=-\delta^{jk}\epsilon^{ilm}n_m+\epsilon^{ilm}n^jn^kn_m=-P^{jk}\epsilon^{ilm}n_m.
\ee

For a periodic system, we may also define the time average of the helicity flux density over a period $T$
\begin{align}
    \langle \frac{dH}{dud\Omega}\rangle=\frac{1}{T}\int_0^T du \frac{dH}{dud\Omega}=\frac{G}{8\pi}\langle \dddot{M}_{ij}\ddot{M}_{kl}\rangle Q^{ijkl}.\label{local2}
\end{align}

As a comparison, we can also reproduce the quadrupole formula for the energy flux density 
\bea 
\frac{dE}{dud\Omega}=-\frac{G}{8\pi}\dddot{M}_{ij}\dddot{M}_{kl}E^{ijkl},
\eea where 
\bea 
E^{ijkl}=\delta^{jl}\delta^{ik}-2\delta^{ik}n^jn^l+\frac{1}{2}n^in^jn^kn^l.
\eea 
Note that the tensor $Q^{ijkl}$ contains odd numbers of $n_i$ in each term and the total helicity flux should be zero 
\begin{align}
\frac{dH}{du}=\int d\Omega \frac{dH}{dud\Omega}=0.
\end{align} 

It seems that there is no non-trivial helicity flux. However, the point is that the expression \eqref{local} is local, and it is expected to detect a non-trivial helicity flux distribution on the celestial sphere. 
In \cite{Seraj:2022qyt}, the authors found that the spin precession of a free falling gyroscopic is affected by the gravitational helicity flux density. By measuring the  spin precession of the gyroscope and the distance of the binary system to the observer at a definite angle, one can obtain the helicity flux density.

For further explanation, one can  extract the angle-dependence of the  helicity flux density by decomposing it into spherical harmonic functions
\bea 
O(u,\Omega)=\sum_{\ell=0}^\infty \sum_{m=-\ell}^\ell {O}_{\ell,m}(u)Y^*_{\ell,m}(\Omega),
\eea  where the coefficients can be solved as 
\bea 
O_{\ell,m}(u)=\int d\Omega O(u,\Omega)Y_{\ell,m}(\Omega).\label{integrated}
\eea 
 Equivalently,
the integrated helicity flux  \eqref{Og} would be non-zero for general function $g(\Omega)$.
For example, we can set $g(\Omega)=Y_{\ell,m}(\Omega)$ and find the mode 
\bea 
\mathcal{O}_{\ell,m}=\mathcal{O}_{g=Y_{\ell,m}}=\int du d\Omega O(u,\Omega)Y_{\ell,m}(\Omega)=\int du O_{\ell,m}(u).
\eea
For periodic system, the time average of the previous mode is 
\be 
\langle \mathcal{O}_{\ell,m}\rangle=\frac{1}{T}\int_0^T du O_{\ell,m}(u).\label{inthe}
\ee 
Therefore, we will focus on angle-dependence of the helicity flux density $O(u,\Omega)$, which is equivalent to the integrated helicity flux \eqref{Og}.

\paragraph{Supertranslation frame.} Note that there is an ambiguity for the shear tensor from BMS transformation \footnote{The shear tensor corresponds to the transverse mode of the gravitational waves which is gauge invariant up to large gauge transformations. Supertranslation is a large gauge transformation which is non-trivial in the sense that the corresponding charge could be non-zero.}
\be 
\delta_{\text{BMS}}C_{AB}=-\left(2\nabla_A\nabla_B-\gamma_{AB}\nabla^2\right)f(\Omega),
\ee where the supertranslation function $f(\Omega)$ on the celestial sphere is related to the choice of the frame. In Bondi formalism, by imposing the standard fall-off conditions for the gravitational waves, the shear tensor is fixed up to a supertranslation\cite{Bondi:1962px,1962RSPSA.270..103S,Barnich:2009se}.  This does not affect the Bondi news tensor, and thus the energy flux density is free from this ambiguity. However, the angular momentum flux density is indeed subject to the supertranslation ambiguity \cite{penrose2011republication,newman1966note}. See recent discussions on this topic in \cite{ashtekar2020compact,Chen:2021szm}. 
Similarly, the helicity flux density transforms up to a total derivative
\bea 
\delta_fO(u,\Omega)=\frac{1}{32\pi G}\dot{C}_{AB}\epsilon^{CA}\left(\gamma^{B}_{C}\nabla^2-2\nabla^B\nabla_C\right)f=\frac{d}{du}\left[\frac{1}{32\pi G}{C}_{AB}\epsilon^{CA}(\gamma^{B}_{C}\nabla^2-2\nabla^B\nabla_C)f\right].\label{ambiguity}
\eea 

During a finite time interval $(t_i,t_f)$, the cumulative variation of the helicity flux density is 
\bea 
\Delta_fO(u,\Omega)=\frac{1}{32\pi G}\epsilon^{CA}(\gamma^{B}_{C}\nabla^2-2\nabla^B\nabla_C)f\times \Delta C_{AB}
\eea
where 
\bea 
\Delta C_{AB}=C_{AB}(t_f)-C_{AB}(t_i).
\eea The result depends on the choice of the supertranslation frame $f(\Omega)$ and the difference in the shear tensor between the initial and final time . In the quadrupole limit, this is transformed to the variation of the quadrupole moment using the relation \eqref{Bondishear}
\bea 
\Delta C_{AB}=2G(Y^i_AY^j_B+\frac{1}{2}\gamma_{AB}n^in^j) \Delta\ddot{M}_{ij},\quad \Delta\ddot{M}_{ij}=\ddot M_{ij}(t_f)-\ddot{M}_{ij}(t_i).
\eea 
For a periodic system in the Newtonian gravity, the variation of the quadrupole moment  vanishes since the stars return to the original locations in a single period $T$. Therefore, we conclude that the time average mode \eqref{inthe} is free from  supertranslation ambiguity \eqref{ambiguity} for  a periodic system in the Newtonian limit since the boundary terms cancel out over a period.

However, the previous result breaks down for non-periodic systems. The variation of the quadrupole moment will never be zero. More explicitly, as we will show in \eqref{ddotM}, the second derivative of the quadrupole moment with respect to time depends on the trajectory of the stars.  Interestingly, for hyperbolic orbits,  the second derivative \eqref{ddotM} is an even function of the angle $\psi$ except for the $\ddot{M}_{12}$ component. Therefore,  the non-vanishing cumulative variation of the quadrupole moment between the initial and final time is
\begin{align}
\Delta \ddot{M}_{12}&=\frac{2GM_1M_2}{a(e^2-1)}\sin A(e (\cos 2 A +3)+4 \cos A ),
\end{align} where  $a$ is the semi-major axis and $e$ is the eccentricity of the hyperbolic orbit. The masses of the two stars are $M_1$ and $M_2$, respectively.  The constant  $A$ is the outgoing angle which is determined by the eccentricity through the parameterization \eqref{parameterA}. The finite shift of the shear tensor becomes
\bea 
\Delta C_{AB}=2G(Y_A^1Y_B^2+Y_A^2Y_B^1+\gamma_{AB}n^1 n^2)\Delta\ddot{M}_{12}.
\eea 

In general, the variation of the helicity flux density  $\Delta_f O(u,\Omega)$ in the  formula \eqref{ambiguity} comes from the ambiguity of the choice of supertranslation frame, which can be parameterized by the value of the shear tensor at $u=-\infty$. In the literature, there are two commonly used  gauges to fix the ambiguity. First, in the canonical gauge, the ambiguity is fixed by the condition $C_{AB}(u=-\infty,\Omega)=0$ and it turns out that the Bondi angular momentum at $u=-\infty$ matches the ADM angular momentum only in this gauge \cite{ashtekar1979angular,ashtekar1979energy}.  Second, in the so-called intrinsic gauge \cite{Veneziano:2022zwh}, the angular momentum is calculated in the center-of-mass frame \cite{damour1981radiation,bini2012gravitational} where the supertranslation vanishes $f=0$. As a consequence, the order $\mathcal{O}(G)$ part of the shear tensor $C_{AB}$ \cite{damour2020radiative} leads to an $\mathcal{O}(G^2)$ contribution to the static angular momentum flux \cite{riva2023angular}. This is also exactly the contribution from the zero-energy gravitons based on amplitude computations \cite{manohar2022radiated,di2022angular}. However, the radiative angular momentum flux is always $\mathcal{O}(G^3)$ and it is widely accepted in  discussions of  compact binary coalescence\cite{Blanchet:2013haa,bishop2016extraction}. In our work, we focus on the radiative helicity flux and will choose the center-of-mass frame  to obtain the result at $\mathcal{O}(G^3)$.

\section{Two-body systems}\label{binarysystem}
In this section, we will derive the angular distribution of the helicity flux density for various orbits of two-body systems in the quadrupole limit. More explicitly, we will review the orbits for two-body systems in Newtonian gravity in subsection \ref{setup}. Then we will turn to the circular orbits in subsection \ref{cir}, elliptic orbits in \ref{ellipticorbits}, hyperbolic orbits in \ref{hyperbolicorbits} and parabolic orbits in \ref{para}. 

\subsection{Setup}\label{setup}
We will discuss the two-body system which is firstly studied in \cite{Peters:1963ux}. The masses of the two stars are denoted as $M_1$ and $M_2$ respectively    and the labels 1 and 2 are used to distinguish the objects. 
The orbital trajectories of the two stars are  
\be 
x_{(i)}=x_{(i)}(t),\quad y_{(i)}=y_{(i)}(t),\quad z_{(i)}=0,\quad i=1,2.
\ee 
The action of this two-body system is 
\bea 
S=\int dt \bigg[\frac{1}{2}M_1 (\dot{x}_{(1)}^2+\dot{y}_{(1)}^2)+\frac{1}{2}M_2(\dot{x}_{(2)}^2+\dot{y}_{(2)}^2)+\frac{GM_1M_2}{\sqrt{(x_{(1)}-x_{(2)})^2+(y_{(1)}-y_{(2)})^2}}\bigg],
\eea and the equations of motion are as follows 
\bea 
\ddot{x}_{(1)}&=&-\frac{GM_2(x_{(1)}-x_{(2)})}{\left((x_{(1)}-x_{(2)})^2+(y_{(1)}-y_{(2)})^2\right)^{3/2}},\\
\ddot{y}_{(1)}&=&-\frac{GM_2(y_{(1)}-y_{(2)})}{\left((x_{(1)}-x_{(2)})^2+(y_{(1)}-y_{(2)})^2\right)^{3/2}},\\
\ddot{x}_{(2)}&=&\frac{GM_1(x_{(1)}-x_{(2)})}{\left((x_{(1)}-x_{(2)})^2+(y_{(1)}-y_{(2)})^2\right)^{3/2}},\\
\ddot{y}_{(2)}&=&\frac{GM_1(y_{(1)}-y_{(2)})}{\left((x_{(1)}-x_{(2)})^2+(y_{(1)}-y_{(2)})^2\right)^{3/2}}.
\eea 
As a consequence, we find 
\bea 
M_1\ddot{x}_{(1)}+M_2 \ddot{x}_{(2)}=M_1\ddot{y}_{(1)}+M_2 \ddot{y}_{(2)}=0.
\eea 
The two stars move around their center-of-mass which may be chosen as the origin of the Cartesian coordinate system 
\be 
M_1 x_{(1)}+M_2 x_{(2)}=M_1 y_{(1)}+M_2y_{(2)}=0.
\ee The distance $D$ of the two stars is 
\be D=\sqrt{(x_{(1)}-x_{(2)})^2+(y_{(1)}-y_{(2)})^2}.
\ee Similarly, the distance between star $i$ ($i=1,2$)  and the center-of-mass would be 
\bea 
D_i=\sqrt{x_{(i)}^2+y_{(i)}^2}.
\eea Therefore, the relation among $D_1, D_2$ and $D$ is 
\be 
D_1=\frac{M_2}{M_1+M_2}D,\quad D_2=\frac{M_1}{M_1+M_2}D,\quad M_1D_1=M_2D_2.
\ee 
We may define a new coordinate system
\bea 
x=x_{(1)}-x_{(2)},\quad y=y_{(1)}-y_{(2)},\quad z=z_{(1)}-z_{(2)},
\eea and then the action becomes
\bea 
S=\int dt \bigg[\frac{1}{2}\frac{M_1M_2}{M_1+M_2}(\dot x^2+\dot y^2)+\frac{GM_1M_2}{\sqrt{x^2+y^2}}\bigg].
\eea 
Therefore, the motion of the two-body system is equivalent to a test particle with a reduced mass
\be 
\mu=\frac{M_1M_2}{M_1+M_2}
\ee moving around a fixed object whose mass is 
\be 
\bar M=M_1+M_2.
\ee The Cartesian coordinates $(x,y)$ may be transformed to the polar coordinates $(D,\psi)$ through 
\bea 
x=D\cos\psi,\quad y=D\sin\psi.
\eea There are several typical orbits which may be parameterized as follows.
\begin{enumerate}
\item Circular orbits. The two stars are separated by a constant distance  
\be 
D=a
\ee  and they revolve around each other at a constant angular velocity 
\bea 
\dot\psi=\sqrt{\frac{G\bar M}{a^3}}.
\eea The period of the orbit is 
\bea 
T=\frac{2\pi}{\dot\psi}=2\pi\sqrt{\frac{a^3}{G\bar M}}.
\eea 

    \item Elliptic orbits. The semi-major axis and the eccentricity of the ellipse are $a$ and $e$ ($0<e<1$), respectively.  Then the elliptic orbit can be parameterized  as 
    \bea 
    D=\frac{\epsilon}{1+e\cos\psi},\quad \epsilon=a(1-e^2).
    \eea 
    The periastron ($\psi=0$) and apoastron ($\psi=\pi$) distances are 
    \bea 
    D_p=\frac{\epsilon}{1+e}=a(1-e),\quad D_a=\frac{\epsilon}{1-e}=a(1+e),
    \eea 
     and the time evolutions of $D$ and $\psi$ are respectively
    \bea 
    \dot D=\sqrt{\frac{G\bar M}{\epsilon}}e\sin\psi,\quad \dot{\psi}=\frac{\sqrt{G\bar M\epsilon}}{D^2}.\label{evolution}
    \eea

   The period of the orbit is 
    \be 
    T=\int_0^{2\pi}\frac{d\psi}{\dot\psi}=2\pi \sqrt{\frac{a^3}{G\bar M}},
    \ee which is formally the same as the period of the circular orbits.
    After some algebra, we find the following conserved energy
\bea 
\frac{1}{2}\mu\dot{D}^2-\frac{G \bar M\mu}{D}=-\frac{GM_1M_2}{2a},
\eea which is indeed negative for any elliptic orbits.
\item Parabolic orbits. The orbital trajectory is represented by 
\be 
D=\frac{\epsilon}{1+\cos\psi}
\ee with eccentricity $e=1$. The orbit can be obtained from elliptic orbits by taking the limit $e\to 1$ while keeping $\epsilon$ finite. This is an unbound orbit and the periastron distance is 
\bea 
D_p=\frac{\epsilon}{2}.
\eea The time evolution of the orbit is the same as \eqref{evolution} with $e=1$ and the initial/final angle $\psi_{\text{in}/\text{out}}$ is
\bea 
\psi_{\text{in}}=-\pi,\quad \psi_{\text{out}}=\pi.
\eea 
\item Hyperbolic orbits. The orbital trajectory may be parameterized by 
\bea 
D=\frac{\epsilon}{1+e\cos\psi},\quad \epsilon=a(e^2-1),
\eea where $a$ is the semi-major axis and the eccentricity $e$ is larger than 1 ($e>1$). The semi-major axis $a$ is related to the initial velocity $v_{\text{in}}$ 
\bea 
a=\frac{G\bar M}{v_{\text{in}}^2},
\eea while the eccentricity may be expressed as  \bea 
e=\sqrt{1+\frac{b^2v_{\text{in}}^4}{G^2\bar M^2}}
\eea  with $b$ the impact parameter. This is an unbound orbit and the initial/final angle $\psi_{\text{in}/\text{out}}$ has been chosen as 
\bea \label{parameterA}
\psi_{\text{in}}=-\arccos\left(-\frac{1}{e}\right)\equiv-A,\quad \psi_{\text{out}}=\arccos\left(-\frac{1}{e}\right)\equiv A.
\eea 
The periastron distance is 
\be 
D_p=\frac{\epsilon}{1+e}=a(e-1),
\ee 
while the time evolution of the distance $D$ and the angle $\psi$ is still given by \eqref{evolution}, except that one should replace $\epsilon $ by $\epsilon=a(e^2-1)$. Note that the parabolic orbits can also be obtained by taking the limit $e\to 1$ while keeping $\epsilon$ finite from hyperbolic orbits.
\end{enumerate}

\subsection{Circular orbits}\label{cir}
As an illustration, we will apply our formula to a binary system where two stars have equal mass $M$. This system can be found in the textbook \cite{carroll2019spacetime}. The two stars are in a circular orbit in the $x$-$y$ plane and their distance is $D=2R$. The radius $R$ is extremely large such that the inner structure of the two stars can be ignored and we will treat them as two points. In the Newtonian limit, the angular frequency of the circular orbit is 
\begin{align} 
\omega=\sqrt{\frac{GM}{4R^3}}.
\end{align} 
The orbits of the two stars are 
\begin{align}
   & x_{(1)}=R \cos\omega t,\quad y_{(1)}=R\sin\omega t,\quad z_{(1)}=0,\\
    & x_{(2)}=-R \cos\omega t,\quad y_{(2)}=-R\sin\omega t,\quad z_{(2)}=0,
\end{align} and the energy density of the system is 
\bea 
T_{00}(t,\bm x)=M\delta(z)[\delta(x-x_{(1)})\delta(y-y_{(1)})+\delta(x-x_{(2)})\delta(y-y_{(2)})].
\eea Therefore, the non-vanishing quadrupole of the binary system is 
\bea 
I_{11}=MR^2(1+\cos2\omega u),\quad I_{22}=MR^2(1-\cos2\omega u),\quad I_{12}=MR^2\sin 2\omega u.
\eea The reduced quadrupole moment becomes 
\bea 
&&M_{11}=\frac{1}{3}MR^2(1+3\cos2\omega u),\\ &&M_{22}=\frac{1}{3}MR^2(1-3\cos2\omega u),\\  &&M_{33}=-\frac{2}{3}MR^2,\\ && M_{12}=M_{21}=MR^2\sin 2\omega u.
\eea 
We emphasise that  $\dddot{M}_{ij}$ is not proportional to $\ddot{M}_{ij}$, one cannot find a vanishing result for the helicity flux density just by the tensor structures of the formula \eqref{local}.
By calculating the second and third time derivative of the reduced quadrupole moment and substituting them into \eqref{local2}, we find the time average of the helicity flux density
\begin{align}
   \langle \frac{dH}{dud\Omega}\rangle&=\frac{G M^2 R^4 \omega ^5 (7 \cos \theta +\cos 3 \theta )}{\pi }\nn\\&=\frac{G^{7/2}M^{9/2}(7 \cos \theta +\cos 3 \theta )}{32 \pi R^{7/2}c^5 },
\end{align} where we have inserted the velocity of light into the formula in the last step. 
Since the period $T=\frac{2\pi}{\omega}\propto G^{-1/2}$, the radiative helicity flux density is indeed $\mathcal{O}(G^3)$ as we estimate in previous section.
Note that the velocity of the star 1 (or 2) is 
\be 
v=\sqrt{\frac{GM}{4R}}, 
\ee we may rewrite the average helicity flux density as 
\bea 
\langle \frac{dH}{du d\Omega}\rangle=\frac{4}{\pi}Mc^2 \left(\frac{v}{c}\right)^7(7 \cos \theta +\cos 3 \theta )={\frac{16}{\pi}Mc^2 \left(\frac{v}{c}\right)^7 \cos\theta(1+\cos^2\theta)}.\label{discir}
\eea
Obviously, it has the dimension of energy since the dimension of the helicity is the same as the angular momentum. We define the characteristic value 
\bea 
E_c=\frac{32}{\pi}Mc^2\left(\frac{v}{c}\right)^7=\frac{G^{7/2}M^{9/2}}{4\pi R^{7/2}c^5}\label{Ecx}
\eea to represent the magnitude of the helicity flux density.

\begin{figure}
    \centering
    \includegraphics{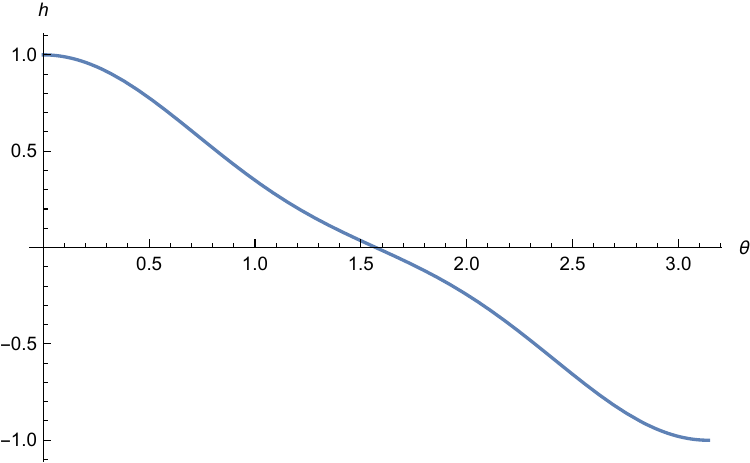}
    \caption{Angle-dependence of the helicity flux. The function $h(\theta)$ is abbreviated to $h$.}
    \label{circular}
\end{figure}

Now we will discuss the angle-dependence of the helicity flux
\be 
h(\theta)=\frac{1}{2}\cos \theta(1+\cos^2\theta)
\ee 
which is shown in figure \ref{circular}.  Since the distribution is invariant under the rotation around the $z$ axis, we only draw the $\theta$ dependence in this figure. 
 Some properties of the angle-dependence function $h(\theta)$ are listed in the following. 
\begin{enumerate}
    \item The helicity flux density is symmetric with respect to the equatorial plane. In other words, it has odd parity  under transformation $\theta\to \pi-\theta$
    \be 
    h(\pi-\theta)=-h(\theta).
    \ee 
    \item Since the function $h(\theta)$ is parity odd, it vanishes at $\theta=\frac{\pi}{2}$
    \be 
    h(\frac{\pi}{2})=0.
    \ee 
    \item The helicity flux density approaches its maximum value at the north pole ($\theta=0$) and minimum value at the south  pole ($\theta=\pi$)
    \bea 
\langle \frac{dH}{du d\Omega}\rangle_{\text{max}}=E_c,\quad \langle\frac{dH}{du d\Omega}\rangle_{\text{min}}=-E_c.  
    \eea  
    \item The function $h(\theta)$ is a monotonic decreasing function 
    \bea 
    h'(\theta)<0,\quad \theta\in (0,\pi).
    \eea 
    \item The angular distribution \eqref{discir} can be transformed to the gauge-invariant integrated helicity flux  $\langle \mathcal{O}_{\ell,m}\rangle $ which is defined in \eqref{inthe}. There are only two non-vanishing modes for circular orbits
\bs\label{modecircle}\begin{align}
\langle\mathcal{O}_{1,0}\rangle&=\frac{8}{5}\sqrt{\frac{\pi}{3}}E_c,\\
\langle\mathcal{O}_{3,0}\rangle&=\frac{2}{5}\sqrt{\frac{\pi}{7}}E_c.
\end{align}\es
The fact that the non-trivial modes have a vanishing magnetic quantum number $m=0$ follows from the axial symmetry of the circular orbit. 

\end{enumerate}

\subsection{Elliptic orbits}\label{ellipticorbits}
\begin{figure}
     \centering
     \includegraphics{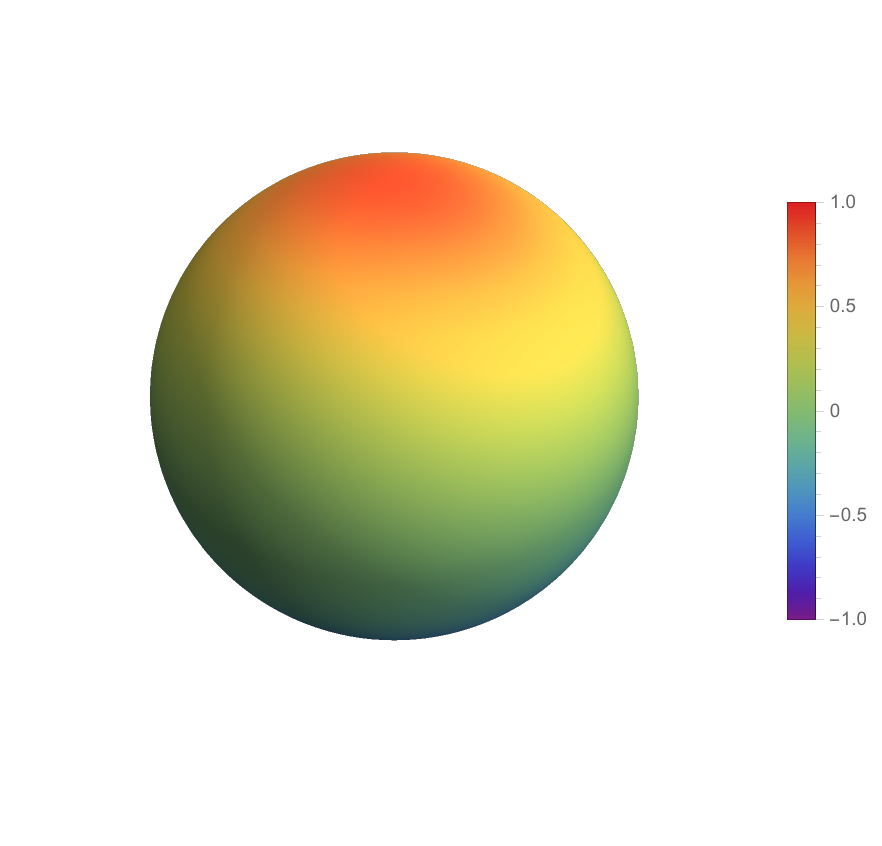}
     \caption{The function $h_e(\theta,\phi)$ on the sphere. We use different color to represent the value of the function. It is obvious that the red color is distributed in the north pole while the blue color in the south pole. }
     \label{gfunc}
 \end{figure}

The orbital trajectories of star 1 and 2 are given by 
\bea 
&& x_{(1)}=D_1\cos\psi,\quad y_{(1)}=D_1\sin\psi,\quad z_{(1)}=0,\\
&& x_{(2)}=-D_2\cos\psi,\quad y_{(2)}={ -}D_2\sin\psi,\quad z_{(2)}=0.
\eea The energy density of the binary system can be expressed as
\bea 
T_{00}(t,\bm x)=M_1\delta(x-x_{(1)})\delta(y-y_{(1)})\delta(z)+M_2\delta(x-x_{(2)})\delta(y-y_{(2)})\delta(z).
\eea Correspondingly, the non-vanishing quadrupole components are
\bea 
I_{11}=I\cos^2\psi,\quad I_{22}=I\sin^2\psi,\quad I_{12}=I_{21}=I\sin\psi\cos\psi,
\eea where $I$ is the trace of the quadrupole 
\bea 
I=\mu D^2.
\eea The reduced quadrupole is 
\bea 
M_{ij}=I\left(\begin{array}{ccc}\cos^2\psi-\frac{1}{3}& \sin\psi\cos\psi&0\\ \sin\psi\cos\psi&\sin^2\psi-\frac{1}{3}&0\\0&0&-\frac{1}{3}\end{array}\right).\label{redquad}
\eea As a consequence, we find 
 \bs\label{ddotM}\begin{align}
 \ddot{M}_{11}&=\frac{G\bar M\mu}{\epsilon}[-\frac{2 e^2}{3}+\frac{1}{6} e (-13 \cos \psi -3 \cos 3 \psi )-2 \cos 2 \psi ],\\
 \ddot{M}_{22}&=\frac{G\bar  M\mu}{\epsilon}[\frac{4 e^2}{3}+\frac{1}{6} e (17 \cos \psi +3 \cos 3 \psi )+2 \cos 2 \psi ],\\
 \ddot{M}_{33}&=-\frac{G\bar M\mu}{\epsilon}\frac{2}{3} e (e+\cos \psi ),\\
 \ddot{M}_{12}&=\ddot{M}_{21}=-\frac{G\bar M\mu}{\epsilon}\sin \psi  [-(e (\cos 2 \psi +3)+4 \cos \psi )]
 \end{align}\es for the second time derivative of the reduced quadrupole and 
\bs\begin{align}
 \dddot{M}_{11}&=\frac{(G\bar M)^{3/2}\mu}{\epsilon^{5/2}}(1+e\cos\psi)^2 \frac{1}{3} \sin \psi  (e (9 \cos 2 \psi +11)+24 \cos \psi ),\\
 \dddot{M}_{22}&=-\frac{(G\bar M)^{3/2}\mu}{\epsilon^{5/2}}(1+e\cos\psi)^2 \frac{1}{3} \sin \psi  (e (9 \cos 2 \psi +13)+24 \cos \psi ),\\
 \dddot{M}_{33}&=\frac{(G\bar M)^{3/2}\mu}{\epsilon^{5/2}}(1+e\cos\psi)^2\frac{2}{3} e \sin \psi ,\\
 \dddot{M}_{12}&=\dddot{M}_{21}=\frac{(G\bar M)^{3/2}\mu}{\epsilon^{5/2}}(1+e\cos\psi)^2\frac{1}{2} (-5 e \cos \psi -3 e \cos 3 \psi -8 \cos 2 \psi )
 \end{align}\es for the third time derivative of the reduced quadrupole.
 We can define a time average quantity
 \be
 G_{ij,kl}=\langle \dddot{M}_{\hspace{-2pt}ij}\ddot{M}_{kl}\rangle=\frac{1}{T}\int_0^T du \dddot{M}_{\hspace{-2pt}ij}\ddot{M}_{kl}=\frac{1}{T}\int_0^{2\pi}d\psi \dddot{M}_{\hspace{-2pt}ij}\ddot{M}_{kl}\dot\psi^{-1}=G_0 g_{ij,kl},
 \ee 
 where $G_0$ is 
 \bea 
 G_0=\frac{(G\bar M)^{5/2}\mu^2}{a^{3/2}\epsilon^2}
 \eea 
 and the  non-vanishing  components of $g_{ij,kl}$ are 
 \bea 
 g_{11,12}&=&g_{11,21}=\frac{1}{12} \left(-37 e^2-48\right),\\
 g_{12,11}&=&g_{21,11}=\frac{1}{12} \left(37 e^2+48\right),\\
 g_{12,22}&=&g_{21,22}=\frac{1}{12} \left(-47 e^2-48\right),\\
 g_{12,33}&=&g_{21,33}=\frac{5 }{6}e^2,\\
 g_{22,12}&=&g_{22,21}=\frac{1}{12} \left(47 e^2+48\right),\\
 g_{33,12}&=&g_{33,21}=-\frac{5}{6} e^2.
 \eea Substituting the results into \eqref{local2}, the average helicity flux density is 
 \bea 
 \langle \frac{dH}{dud\Omega}\rangle&=&\frac{1}{8\pi}\frac{G^{7/2}\bar M^{5/2}\mu^2}{a^{3/2}\epsilon^2}g_{ij,kl}Q^{ij,kl}.\eea 
 More explicitly, \bea \langle \frac{dH}{dud\Omega}\rangle&=&\frac{1}{4\pi}\frac{G^{7/2}\bar M^{5/2}\mu^2}{a^{3/2}\epsilon^2}\left[(7 \cos \theta +\cos 3 \theta )+\frac{1}{4} e^2 \cos \theta  \left(5 \sin ^2\theta  \cos 2 \phi +7 \cos 2 \theta +21\right)\right].\nn\\\label{helicityelliptic}
 \eea The result is consistent with the one in circular orbit studied in previous subsection\footnote{One should set $M_1=M_2=M,\bar M=2M,\mu=\frac{M}{2},a=2R,e=0$}.
 As the circular case, we may define the angle-dependence of the helicity flux as 
 \bea 
 h(\theta,\phi;e)=\frac{1}{8}\left[(7 \cos \theta +\cos 3 \theta )+\frac{1}{4} e^2 \cos \theta  \left(5 \sin ^2\theta  \cos 2 \phi +7 \cos 2 \theta +21\right)\right].\label{he}
 \eea The function $h(\theta,\phi;e)$ depends on the spherical angles and the eccentricity of the elliptic orbit. Taking the limit as $e\to 0$, it is the same as the function $h(\theta)$. Therefore, it would be better to focus on the contribution from the eccentricity and define a new function 
 \be 
 h_e(\theta,\phi)=\frac{1}{32}\cos \theta  \left(5 \sin ^2\theta  \cos 2 \phi +7 \cos 2 \theta +21\right),
 \ee whose properties are shown in the following.
 \begin{enumerate}
     \item Discrete symmetry 
     \bea 
     h_e(\pi-\theta,\phi)=-h_e(\theta,\phi),\quad h_e(\theta,\pi\pm\phi)=h_e(\theta,\phi).
     \eea Therefore, the function $h_e(\theta,\phi)$ is still parity odd 
     \be 
     h_e(\pi-\theta,\pi+\phi)=-h_e(\theta,\phi).
     \ee 
    \item The rotation symmetry around the $z$ axis is broken and the function $h_e(\theta,\phi)$ depends on $\phi$ explicitly. It is not hard to show that the maximum value of $h_e(\theta,\phi)$ locates at \bea 
    (\theta,\phi)=(0,0) \quad \text{or}\quad (0,\pi)
    \eea with
    \bea 
    h_e(0,0)=h_e(0,\pi)=\frac{7}{8}.
    \eea By the parity transformation, we find that the minimum value of $h_e(\theta,\phi)$ locates at
    \be 
    (\theta,\phi)=(\pi,0)\quad\text{or}\quad (\pi,\pi)
    \ee with
    \be 
    h_e(\pi,0)=h_e(\pi,\pi)=-\frac{7}{8}.
    \ee 
 \end{enumerate} We draw the function $h_e(\theta,\phi)$ on the sphere in figure \ref{gfunc} where 
 the angle-dependence of the helicity flux is depicted by color in the figure. 

\begin{figure}[htbp]
    \centering
    \begin{subfigure}[t]{0.45\linewidth}
        \centering
        \includegraphics[width=\linewidth]{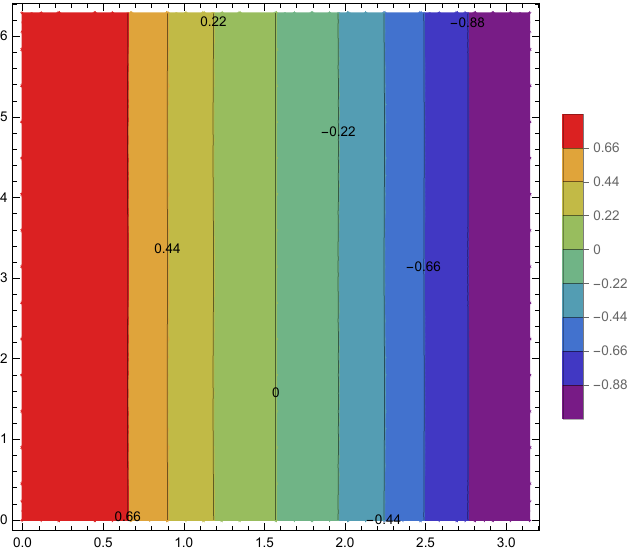}
        \caption{
        }
        \label{e02}
    \end{subfigure}
    \hspace{0.05\linewidth}
    \begin{subfigure}[t]{0.45\linewidth}
        \centering
        \includegraphics[width=\linewidth]{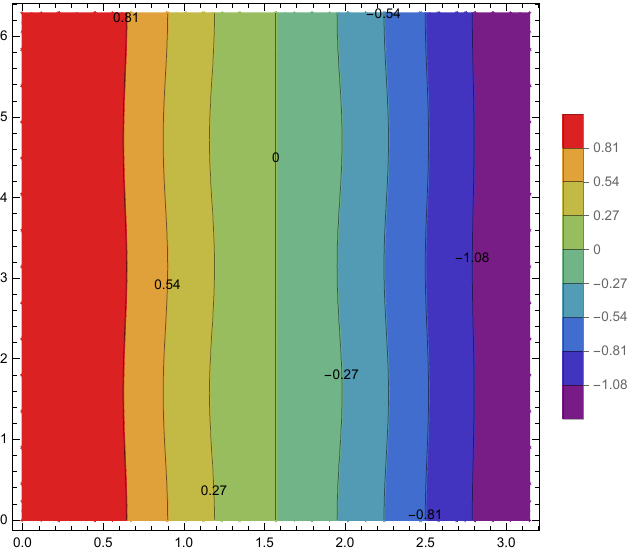}
        \caption{
        }
        \label{e05}
    \end{subfigure}
    \\
    \begin{subfigure}[t]{0.6\linewidth}
    \centering
    \includegraphics[scale=0.8]{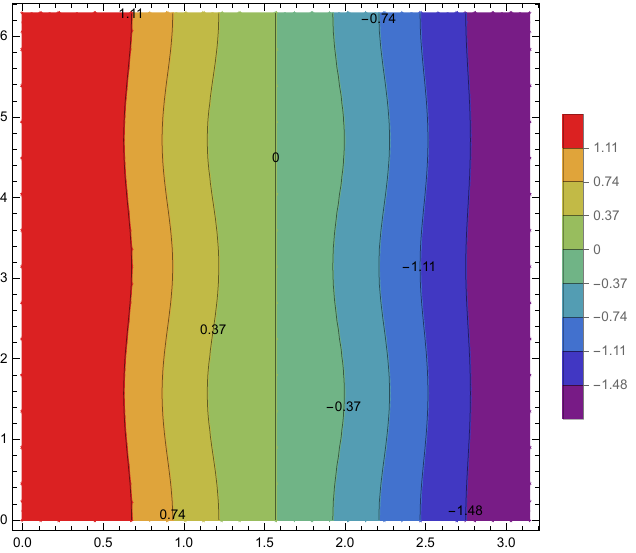}
    \caption{
    }
    \label{e09}
    \end{subfigure}
    \caption{The contour map for the function $h(\theta,\phi;e)$ for different values of eccentricity $e$. The eccentricity is $e=0.2, 0.5, 0.9$ for figure a, b, c respectively. The numbers on the contour lines are the values of the function $h(\theta,\phi;e)$ for the corresponding lines. The absolute value of $h(\theta,\phi;e)$ increases monotonously with increasing eccentricity.}
\end{figure}

\begin{figure}[htbp]
    \centering
    \begin{subfigure}[t]{0.45\linewidth}
        \centering
        \includegraphics[width=\linewidth]{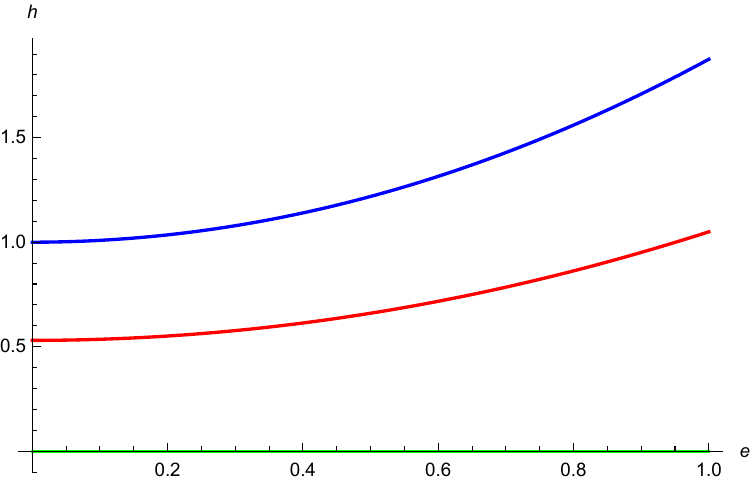}
        \caption{
        }
        \label{helicitye}
    \end{subfigure}
    \hspace{0.05\linewidth}
    \begin{subfigure}[t]{0.45\linewidth}  
        \centering
        \includegraphics[width=\linewidth]{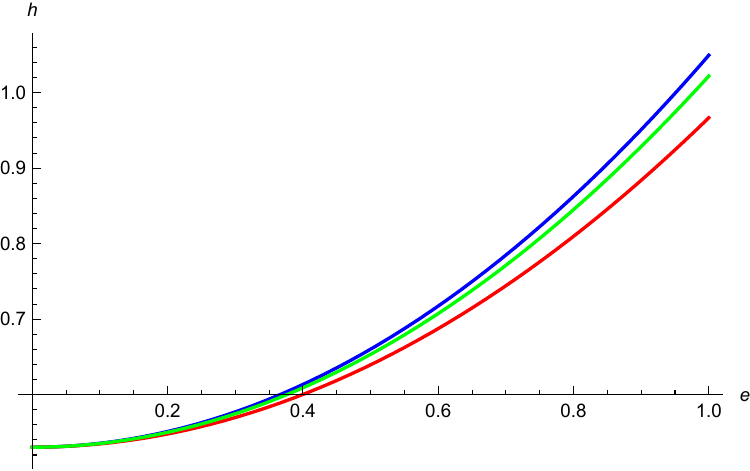} 
        \caption{
        }
        \label{helicityphi}
    \end{subfigure}
    \caption{The dependence on eccentricity of the function $h(\theta,\phi;e)$ at fixed spherical angles. In figure a, the azimuthal angle is fixed to $\phi=0$ and the polar angle $\theta=0,\frac{\pi}{4},\frac{\pi}{2}$ for blue, red and green curves correspondingly. In figure b, the  polar angle is fixed to $\theta=\frac{\pi}{4}$ and the azimuthal angle $\phi=0,\frac{\pi}{3},\frac{5\pi}{6} $ for  blue, red and green curves correspondingly.}
\end{figure}

We may also draw a contour map in the $\theta$-$\phi $ plane for the function $h(\theta,\phi;e)$.  
Since the function $h(\theta, \phi;e)$ also depends on the eccentricity, we can choose different values of $e$. From figure \ref{e02} to figure \ref{e09},  we have set $e=0.2, 0.5, 0.9$ respectively and draw the contour map for the function $h(\theta,\phi;e)$. The dependence of the axis angle $\phi$ becomes more and more important as $e$ increases. To emphasize the dependence on the eccentricity, we also draw the function $h(\theta,\phi;e)$ at fixed spherical angles. In figure 
 \ref{helicitye}, we fix $\phi=0$ and show the helicity flux density at $\theta=0,\frac{\pi}{4},\frac{\pi}{2}$, respectively. The helicity flux density also depends on $\phi$. In figure \ref{helicityphi}, we fix $\theta=\frac{\pi}{4}$ and draw the helicity flux density at $\phi=0,\frac{\pi}{3},\frac{5\pi}{6}$, respectively.

Now we can also transform the helicity flux density to the integrated helicity flux whose non-vanishing components are 
\bs\label{modeellip}\begin{align}
\langle \mathcal{O}_{1,0}\rangle&=\frac{1}{5}\left(8+7 e^2\right)\sqrt{\frac{\pi }{3}} \tilde{E}_c,\\
\langle\mathcal{O}_{3,0}\rangle&=\frac{1}{20}\left(8+7 e^2\right)\sqrt{\frac{\pi }{7}} \tilde{E}_c,\\
\langle\mathcal{O}_{3,2}\rangle&=\frac{e^2}{8}\sqrt{\frac{5 \pi }{42}}\tilde{E}_c,\\
\langle\mathcal{O}_{3,-2}\rangle&=\frac{e^2}{8}\sqrt{\frac{5 \pi }{42}}\tilde{E}_c.
\end{align}\es
where 
\be 
\tilde{E}_c=\frac{2}{\pi}\frac{G^{7/2}\bar M^{5/2}\mu^2}{a^{3/2}\epsilon^2}.
\ee Comparing above equations with \eqref{modecircle}, the modes with $\ell=3,m=\pm 2$ are non-zero since the eccentricity $e\not=0$ , reflecting the axial angle-dependence of the helicity flux density for elliptic orbits.

\subsection{Hyperbolic orbits}\label{hyperbolicorbits}
 We find the same reduced quadrupole \eqref{redquad} by replacing $\epsilon=a(1-e^2)$ to $\epsilon=a(e^2-1)$. Since the hyperbolic orbit is unbound, we cannot define the time average over a period. However, we can compute the total helicity flux density by 
 \bea 
 \frac{dH}{d\Omega}=\int du \frac{dH}{du d\Omega}=\frac{G}{8\pi}\int_{\psi_{\text{in}}}^{\psi_{\text{out}}}d\psi \dddot{M}_{ij}\ddot{M}_{kl}Q^{ijkl}\dot{\psi}^{-1}\label{hyperbolic},\eea where 
 the integrand is
 \bea 
 \frac{G}{8\pi}\dddot{M}_{ij}\ddot{M}_{kl}Q^{ijkl}\psi^{-1}=\frac{G^3\bar{M}^2\mu^2}{16\pi\epsilon^2}K(\theta,\phi;\psi)
 \eea with 
 \bea 
 K(\theta,\phi;\psi)&=&(1+e\cos\psi)\cos\theta\times [8 (3+\cos 2 \theta )\nn\\&&+e \left(12 (\cos2 \theta +3) \cos \psi +2 \sin ^2\theta  (3 \cos (2 \phi -\psi )+\cos (2 \phi -3 \psi ))\right)\nn\\&&+e^2 \left((\cos 2 \theta +3) (3 \cos (2 \psi )+1)+2 \sin ^2\theta  (3 \cos (2 (\phi -\psi ))+\cos 2 \phi )\right) ].
 \eea Therefore, the integral \eqref{hyperbolic} is 
 \bea 
 \frac{dH}{d\Omega}=\frac{G^3\bar{M}^2\mu^2}{16\pi\epsilon^2}\cos\theta \kappa(\theta,\phi;A)\label{distru}
 \eea with 
 \bea 
 \kappa(\theta,\phi;A)&=&2A \left(5 e^2 \sin ^2\theta  \cos 2 \phi +\left(7 e^2+8\right) \cos 2 \theta +21 e^2+24\right)\nn\\&&+\frac{1}{3} e \sin A [\sin ^2\theta  \cos 2 \phi  \left(4 \left(3 e^2+2\right) \cos 2 A+63 e \cos A+3 e \cos 3 A+36 e^2+40\right)\nn\\&& +6 (\cos 2 \theta +3) \left(e^2 \cos 2 A+9 e \cos A+3 e^2+20\right)].
 \eea Still, the total helicity flux is zero while its angle-dependence \eqref{distru} is nontrivial. To check the consistency of our result, we continue the result to the elliptic orbits with $0<e<1$. In this case, the integral domain should be  $(-\pi,\pi)$ which corresponds to $A=\pi$. Substituting $A=\pi$ into Eq.\eqref{distru}, we find the following average helicity flux density over a period 
 \bea 
 \langle \frac{dH}{du d\Omega}\rangle&=&\frac{1}{T}\frac{G^3\bar{M}^2\mu^2}{16\pi\epsilon^2}\cos\theta \kappa(\theta,\phi;\pi)\nn\\&=&\frac{G^{7/2}\bar{M}^{5/2}\mu^2}{16\pi a^{3/2}\epsilon^2}\cos \theta \left(5 e^2 \sin ^2\theta  \cos 2 \phi +\left(7 e^2+8\right) (\cos 2 \theta +3)\right),
 \eea which is exactly the equation \eqref{helicityelliptic}.
 
\begin{figure}[htbp]
    \centering
    \begin{subfigure}[t]{0.45\linewidth}
        \centering
        \includegraphics[width=\linewidth]{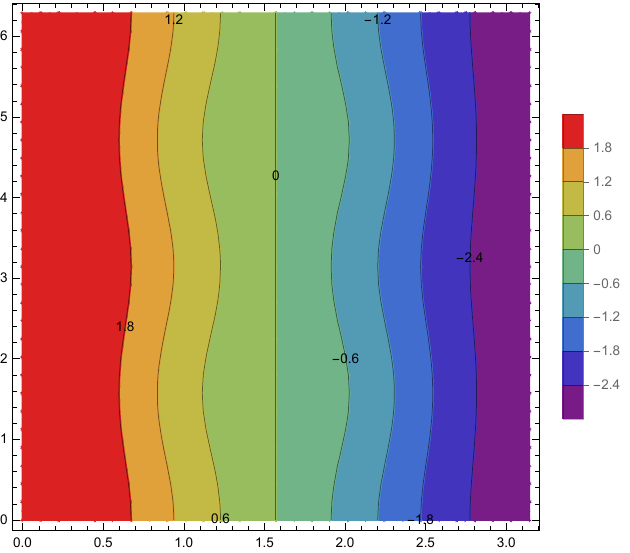}
        \caption{
        }
        \label{e1d4}
    \end{subfigure}
    \hspace{0.05\linewidth} 
    \begin{subfigure}[t]{0.45\linewidth}
        \centering
        \includegraphics[width=\linewidth]{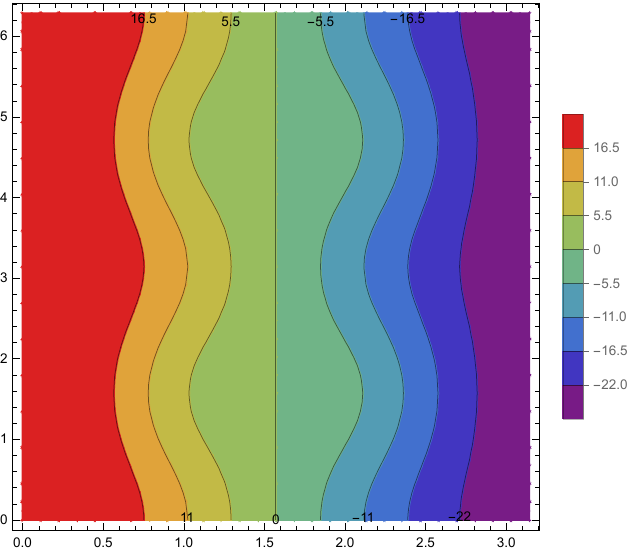} 
        \caption{
        }
        \label{e5}
    \end{subfigure}
    \\
    \begin{subfigure}[t]{0.6\linewidth}
    \centering
    \includegraphics[scale=0.8]{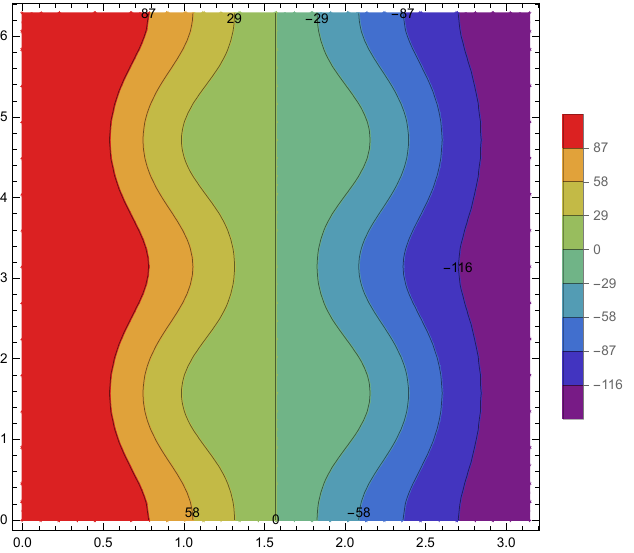}
    \caption{
    }
    \label{e10}
    \end{subfigure}
    \caption{The angle-dependence of the total helicity flux in the $\theta$-$\phi$ plane. The eccentricity is $e=1.4, 5, 10$ for figure a, b, c, respectively. In figure a, the pattern of the contour lines is similar to
the one in figure \ref{e09}. In figure b, the
contour lines are more tortuous than the one in figure a. In figure c, the pattern of the
contour lines are almost the same as ones in figure  b, except that the absolute values become much more larger.}
\end{figure}

 The function $\kappa(\theta,\phi;A)$ may be separated into two parts. The first part depends linearly  on $A$ 
 \bea 
 \kappa_1(\theta,\phi;A)=2A \left(5 e^2 \sin ^2\theta  \cos 2 \phi +\left(7 e^2+8\right) \cos 2 \theta +21 e^2+24\right),
 \eea while the second part is a superposition of sine and cosine functions of $A$
 \bea 
 \kappa_2(\theta,\phi;A)&=&\frac{1}{3} e \sin A [\sin ^2\theta  \cos 2 \phi  \left(4 \left(3 e^2+2\right) \cos 2 A+63 e \cos A+3 e \cos 3 A+36 e^2+40\right)\nn\\&& +6 (\cos 2 \theta +3) \left(e^2 \cos 2 A+9 e \cos A+3 e^2+20\right)]\nn\\&=&\frac{2\sqrt{e^2-1}}{3e^2}[3 \left(2 e^2+13\right) e^2 (\cos 2 \theta +3)+\left(12 e^4+e^2+2\right) \sin ^2\theta  \cos 2 \phi ].
 \eea In the high eccentricity limit, the asymptotic behaviors of $\kappa_1$ and $\kappa_2$
 are 
 \bea 
 \kappa_1(\theta,\phi;A)&\sim&\pi  e^2 \left(5 \sin ^2\theta  \cos 2 \phi +7 \cos 2 \theta +21\right),\\
 \kappa_2(\theta,\phi;A)&\sim&4 e^3 \left(2 \sin ^2\theta  \cos 2 \phi +\cos 2 \theta +3\right).
 \eea Therefore, the second part becomes more important than the first part in the high eccentricity limit. On the other hand, taking the limit as $e\to 1$, the second part becomes zero while the first part is still finite. To find the characteristic value of $e$ such that $\kappa_1\approx \kappa_2$, we notice that both of them reach their maximum values at the north pole 
 \bea 
 \kappa_1(0,0;A)&=&8 \left(7 e^2+8\right) \arccos \left(-\frac{1}{e}\right),\\
 \kappa_2(0,0;A)&=&8 \sqrt{e^2-1} \left(2 e^2+13\right).
 \eea They are equal to each other at 
 \bea 
 e\approx e_c=5.3.
 \eea 
 We separate hyperbolic orbits into two classes according to the eccentricity, 
$1<e <e_c,\ e>e_c$. From figure \ref{e1d4} to figure \ref{e10}, we draw the contour map for the angle-dependence of the total helicity flux for $e=1.4,\ 5$ and $10$. We have renormalized the angle-dependence as 
\bea 
{h}(\theta,\phi;e)=\frac{1}{64\pi}\cos\theta \kappa(\theta,\phi;A)
\eea such that it matches \eqref{he} in the limit $e\to 1$. 

\begin{figure}[htbp]
    \centering
    \begin{subfigure}[t]{0.45\linewidth} 
        \centering
        \includegraphics[width=\linewidth]{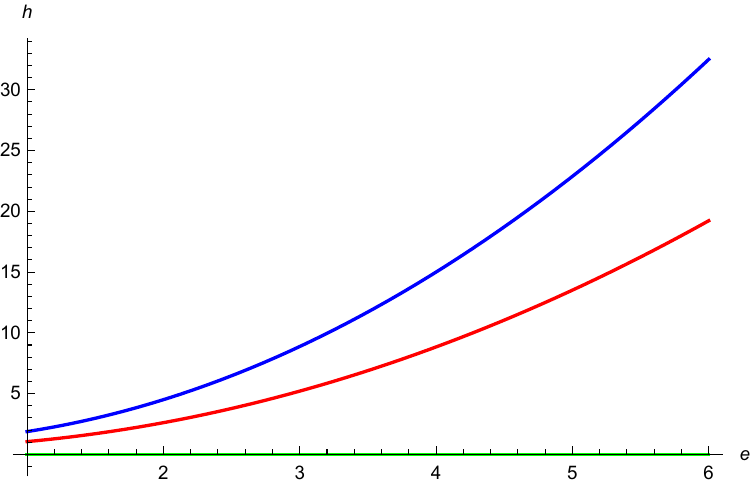}
        \caption{
        }
        \label{helicitye2}
    \end{subfigure}
    \hspace{0.05\linewidth} 
    \begin{subfigure}[t]{0.45\linewidth} 
        \centering
        \includegraphics[width=\linewidth]{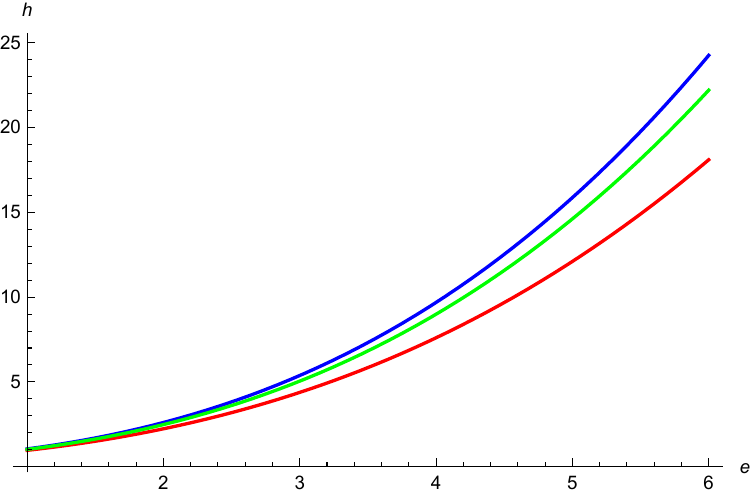}
        \caption{
        }
        \label{helicityphi2}
    \end{subfigure}
    \caption{The dependence on eccentricity with $e\in(1,6)$ of the function $h(\theta,\phi;e)$ at  fixed spherical angles. In figure a, the azimuthal angle is fixed to $\phi=0$ and the polar angle $\theta=0,\frac{\pi}{4},\frac{\pi}{2}$ for blue, red and green curves correspondingly. In figure b, the  polar angle is fixed to $\theta=\frac{\pi}{4}$ and the azimuthal angle $\phi=0,\frac{\pi}{3},\frac{5\pi}{6} $ for  blue, red and green curves correspondingly.}
\end{figure}

Similar to the elliptic orbits, we can also fix the spherical angles and draw the dependence on the eccentricity for the helicity flux density. This is shown in figure \ref{helicitye2} and \ref{helicityphi2}. The qualitative  behaviour is the same as figure \ref{helicitye} and \ref{helicityphi} correspondingly. However, the magnitude becomes much larger compared to the elliptic orbits.

Note that we can also compute the total energy flux density for hyperbolic orbits 
\bea 
\frac{dE}{d\Omega}=-\frac{G}{8\pi}\int_{\psi_{\text{in}}}^{\psi_{\text{out}}} d\psi\dddot{M}_{ij}\dddot{M}_{kl}E^{ijkl}\dot\psi^{-1},
\eea where the integrand is
\bea 
-\frac{G}{8\pi}\dddot{M}_{ij}\dddot{M}_{kl}E^{ijkl}\dot\psi^{-1}=-\frac{G^{7/2}\bar{M}^{5/2}\mu^2}{512\pi\epsilon^{7/2}}p(\theta,\phi;\psi)
\eea with 
\bea 
p(\theta,\phi;\psi)&=&(1+e\cos\psi)^2\left(-2 \sin ^2\theta  \cos 2 (\phi -\psi )+\cos 2 \theta +3\right)\nn\\&&\times [36 e^2 \sin ^2\theta  \sin 4 \psi  \sin 2 \phi +8 \left(15 e^2+32\right) \sin ^2\theta  \sin 2 \psi  \sin 2 \phi +36 e^2 \sin ^2\theta  \cos 4 \psi  \cos 2 \phi \nn\\&&+2 \cos 2 \psi  \left(4 \left(15 e^2+32\right) \sin ^2\theta  \cos 2 \phi +30 e^2 (\cos 2 \theta +3)\right)
-25 e^2 \cos 2 (\theta -\phi )\nn\\&&-25 e^2 \cos 2 (\theta +\phi )+68 e^2 \cos 2 \theta +50 e^2 \cos 2 \phi +204 e^2+320 e \sin ^2\theta  \sin \psi  \sin 2 \phi \nn\\&&+192 e \sin ^2\theta  \sin 3 \psi  \sin 2 \phi +192 e \sin ^2\theta  \cos 3 \psi  \cos 2 \phi \nn\\&&+64 e \cos \psi  \left(5 \sin ^2\theta  \cos 2 \phi +4 \cos 2 \theta +12\right)+128 \cos 2 \theta +384].
\eea Therefore, the total energy flux density is 
\bea 
\frac{dE}{d\Omega}=-\frac{G^{7/2}\bar{M}^{5/2}\mu^2}{512\pi\epsilon^{7/2}}\eta(\theta,\phi;A)
\eea 
with 
\bea 
\eta(\theta,\phi;A)&=&\eta_1(\theta,\phi;A)+\eta_2(\theta,\phi;A).
\eea The first part $\eta_1(\theta,\phi;A)$ is proportional to $A$
\bea 
\eta_1(\theta,\phi;A)&=&A\bigg[64 (28 \cos 2 \theta +\cos 4 \theta +35)+e^2 \left(208 \sin ^2\theta  (\cos 2 \theta +3) \cos 2 \phi +5416 \cos 2 \theta \right.\nn\\&&+\left.198 \cos 4 \theta +6802\right)+e^4 \left(-50 \sin ^4\theta  \cos 4 \phi +32 \sin ^2\theta  (\cos 2 \theta +3) \cos 2 \phi \nn\right.\\&&+\left.682 \cos 2 \theta +\frac{51}{2} \cos 4 \theta +\frac{1721}{2}\right)\bigg],
\eea while the second part is 
\bea 
\eta_2(\theta,\phi;A)&=&\frac{\sqrt{e^2-1}}{30e^4}\bigg[704 \sin ^4\theta  \cos 4 \phi +e^2 \left(-2912 \sin ^4\theta  \cos 4 \phi -320 \sin ^2\theta  (\cos 2 \theta +3) \cos 2 \phi \right)\nn\\&&+e^4 \left(4712 \sin ^4\theta  \cos 4 \phi +2080 \sin ^2\theta  (\cos 2 \theta +3) \cos 2 \phi \right.\nn\\&&+\left.5 (22440 \cos 2 \theta +806 \cos 4 \theta +28082)\right)\bigg].\nn\\
\eea 
Unlike the total helicity flux, the total energy flux is non-vanishing 
\bea 
\Delta E&=&\int d\Omega \frac{dE}{d\Omega}\nn\\&=&-\frac{G^{7/2}\bar{M}^{5/2}\mu^2}{45\epsilon^{7/2}}\bigg[2 \sqrt{e^2-1} \left(673 e^2+602\right)+6 \arccos\left(-\frac{1}{e}\right)\left(37 e^4+292 e^2+96\right)\bigg].\nn\\
\eea The result matches the one in \cite{1977ApJ...216..610T}.
 \subsection{Parabolic orbits}\label{para}
 For parabolic orbits, the total helicity flux density can be found by setting $e=1$ while keeping $\epsilon$ finite. Therefore, we find 
 \bea 
 \frac{dH}{d\Omega}=\frac{5G^3\bar M^2\mu^2}{8\epsilon^2}\cos \theta  \left(\sin ^2\theta  \cos 2 \phi +3 \cos 2 \theta +9\right).
 \eea The maximum value of $\frac{dH}{d\Omega}$ locates at the points 
 \bea 
 (\theta,\phi)=(0,0)\quad\text{or}\quad  (0,\pi)
 \eea with 
 \bea 
 \frac{dH}{d\Omega}\Big|_{\text{max}}=\frac{15G^3\bar M^2\mu^2}{2\epsilon^2}.
 \eea Correspondingly, the minimum value of $\frac{dH}{d\Omega}$ locates at the points 
 \bea 
 (\theta,\phi)=(\pi,0)\quad\text{or}\quad  (\pi,\pi)
 \eea with 
 \be 
  \frac{dH}{d\Omega}\Big|_{\text{min}}=-\frac{15G^3\bar M^2\mu^2}{2\epsilon^2}.
 \ee

\section{Higher multipoles}\label{higher}
In the previous section, we mainly focused on the contribution of the mass quadrupole. In general, there are higher multipoles contributing to the radiative fluxes. Near future null infinity, the symmetric trace free tensor can be expressed as two types of radiative multipole moments \cite{Thorne:1980ru}
\bea 
h_{ij}^{\text{TT}}=\frac{4G}{r}(P_i^{(i'}P_j^{j')}-\frac{1}{2}P_{ij}P^{i'j'})\sum_{\ell=2}^\infty \frac{n^{i(\ell-2)}}{\ell!}\left(U_{i'j'i(\ell-2)}-\frac{2\ell}{\ell+1}\epsilon_{i'pq}n_pV_{j'q i(\ell-2)}\right)+\mathcal{O}(r^{-2}),
\eea where $U_{i(\ell)}$ are mass-type multipole moments and $V_{i(\ell)}$ are current-type multipole moments. Both of them are symmetric trace free and we use the notation $i(\ell)$ to indicate that the $\ell$ indices $i_1,i_2,\cdots,i_\ell$ are symmetric trace free.
 The radiative multipole moments are  functionals of the source canonical moments $M_{i(\ell)}$ and $S_{i(\ell)}$ 
\bea 
U_{i(\ell)}= M^{(\ell)}_{i(\ell)}+\mathcal{O}(G),\quad V_{i(\ell)}=S^{(\ell)}_{i(\ell)}+\mathcal{O}(G)\label{corrections}
\eea with 
\bea 
M^{(\ell)}_{i(\ell)}\equiv\frac{d^\ell}{du^\ell}M_{i(\ell)},\quad S^{(\ell)}_{i(\ell)}\equiv \frac{d^\ell}{du^\ell}S_{i(\ell)}.
\eea The higher order post-Newtonian corrections to the radiative multipole moments are known and we refer the reader to \cite{Blanchet:2013haa}. The shear tensor may be found as 
\bea 
C_{AB}=4G P_{AB}^{ij}\sum_{\ell=2}^\infty \frac{n^{i(\ell-2)}}{\ell!}\left(U_{iji(\ell-2)}-\frac{2\ell}{\ell+1}\epsilon_{ipq}n_pV_{jq i(\ell-2)}\right),
\eea where 
\bea 
P_{AB}^{ij}=Y^{i'}_A Y^{j'}_B (P^i_{(i'}P^j_{j')}-\frac{1}{2}P^{ij}P_{i'j'})=Y_A^{(i}Y_B^{j)}-\frac{1}{2}\gamma_{AB}P^{ij}.
\eea 
Correspondingly, the helicity flux density is 
\bea 
O(u,\Omega)&=&\frac{1}{32\pi G}\dot{C}_{AB}C^B_{\ C}\epsilon^{CA}\nn\\&=&
\frac{G}{2\pi}P^{iji'j'}\sum_{\ell,\ell'=2}^\infty \frac{n^{i(\ell-2)}}{\ell!}\frac{n^{i'(\ell'-2)}}{\ell'!}\left(\dot{U}_{iji(\ell-2)}-\frac{2\ell}{\ell+1}\epsilon_{ipq}n_p\dot{V}_{jq i(\ell-2)}\right)\nn\\&&\times\left(U_{i'j'i'(\ell'-2)}-\frac{2\ell'}{\ell'+1}\epsilon_{i'p'q'}n_{p'}V_{j'q' i'(\ell'-2)}\right),\nn\\
\eea 
where the rank 4 tensor $P^{iji'j'}$ is 
\bea 
&&P^{iji'j'}=P^{ij}_{AB}P^{i'j'B}_{\hspace{0.7cm}C}\epsilon^{CA}=-\frac{1}{4}(\epsilon^{ij'm}P^{ji'}+\epsilon^{ii'm}P^{jj'}+\epsilon^{ji'm}P^{ij'}+\epsilon^{jj'm}P^{ii'})n_m.
\eea 
The angle-dependence of the helicity flux is given by the previous formula. As a consequence, we may find the total radiative rate of the  helicity flux
\bea 
\frac{dH}{du}=\int d\Omega \frac{dH}{du d\Omega}=\int d\Omega O(u,\Omega)=\frac{G}{2\pi}\sum_{\ell=2}^\infty \frac{\ell+2}{(2\ell+1)!! \ell!(\ell-1)}[\dot{U}_{i(\ell)}V_{i(\ell)}-\dot{V}_{i(\ell)}U_{i(\ell)}].\label{total}
\eea 
We have used the integrals of the product of the symmetric trace free tensors on the unit sphere which can be found in the Appendix \ref{integrals}.
As a consistency check, we also compute the energy flux density operator 
\bea 
\hspace{-0.3cm}T(u,\Omega)&=&\frac{G}{4\pi}(P^{ii'}P^{jj'}+P^{ij'}P^{ji'}-P^{ij}P^{i'j'})\nn\\&&\times \sum_{\ell,\ell'=2}^\infty \frac{n^{i(\ell-2)}}{\ell!}\frac{n^{i'(\ell'-2)}}{\ell'!}\left(\dot{U}_{iji(\ell-2)}-\frac{2\ell}{\ell+1}\epsilon_{ipq}n_p \dot{V}_{jq i(\ell-2)}\right)\nn\\
&&\times\left(\dot{U}_{i'j'i'(\ell'-2)}-\frac{2\ell'}{\ell'+1}\epsilon_{i'p'q'}n_{p'}\dot{V}_{j'q' i'(\ell'-2)}\right).\nn\\
\eea Therefore, the total energy flux is 
\bea 
\frac{dE}{du}&=&-\int d\Omega\, T(u,\Omega)\nn\\&=&-\frac{G}{4\pi}\sum_{\ell=2}^\infty\frac{(\ell+1)(\ell+2)}{(2\ell+1)!!\ell! \ell(\ell-1)}[\dot{U}_{i(\ell)}\dot{U}_{i(\ell)}+\left(\frac{2\ell}{\ell+1}\right)^2\dot{V}_{i(\ell)}\dot{V}_{i(\ell)}],
\eea which is exactly the one in \cite{Blanchet:2013haa}.

\section{Application}\label{application}
In the asymptotic flat region of binary black hole systems, the gravitational field is weak and we may apply the previous results to these real systems.  
We can estimate the magnitude of the helicity flux for the event GW150914  as follows. The mass of the two black holes is approximately the same
\be 
M\approx 30 M_{\odot}
\ee and their distance is estimated as
\be 
D=350\text{km}.
\ee Therefore, the characteristic magnitude of the helicity flux density is 
\be 
E_c\approx 3.5\times 10^{45}\text{kg}\cdot \text{m}^2/\text{s}^2\approx 3\times 10^{79} \hbar/\text{s}.
\ee There is a huge number of gravitons radiated out. However, due to the large distance $d$ of the event\footnote{The luminosity distance of the event GW150914   is approximately $d_{\text{L}}\approx 410\text{Mpc}$ \cite{LIGOScientific:2016aoc} .}, the gravitons are diluted when arriving at the earth. Note that the distance in an expanding universe is rather tricky \cite{dodelson2020modern}. The comoving distance $d_{\text{c}}$ is always fixed as the universe expands while the luminosity distance  $d_{\text{L}}$ is  defined as
\bea 
d_{\text{L}}=\sqrt{\frac{L}{4\pi F}},\label{dL}
\eea where $L$ is the luminosity of the star and $F$ is the measured energy flux from the object. It is related to the transverse comoving distance $d_{\text{c}}$ through 
\be 
d_{\text{L}}=(1+z)d_{\text{c}},\label{dLdc}
\ee where $z$ is the redshift factor. Another useful distance in astronomy is the angular diameter distance $d_{\text{A}}$ which is related to comoving distance through
\be 
d_{\text{A}}=\frac{d_{\text{c}}}{1+z}.
\ee

Since the redshift factor $z\approx 0.09$ for the event GW150914, we may just 
use the luminosity distance of the event GW150914 to estimate the magnitude.

Actually, since the helicity flux is independent  of the energy flux, we may use it to define a new distance 
\bea 
d_{\text{h}}=\sqrt{\frac{O}{4\pi \dot S}}.\label{dh}
\eea where $\dot S$ is the non-linear term in gyroscopic spin precession caused by radiative helicity flux \cite{Seraj:2022qyt}.
The formula \eqref{dh} is the analog of the luminosity distance \eqref{dL}. The helicity flux density counts the number density difference of the gravitons with left and right helicity per unit time. The number of the gravitons across the spherical shell is invariant. However, due to the expansion of the universe, the helicity flux density becomes smaller by a factor $a=\frac{1}{1+z}$. Therefore, the new distance $d_{\text{h}}$ is related to the comoving distance through 
\bea 
d_{\text{h}}=\sqrt{1+z}d_{\text{c}}.
\eea 
In comparison with \eqref{dLdc}, there is a $\sqrt{1+z}$ discrepancy since the frequency of the photon is redshifted  while the particle number is not. 
It may provide an independent way to measure the cosmological distance and contribute to the resolution of Hubble tension.

The radius of the orbital system is decreasing due to the radiation of the energy. In a period, the radiative energy is  
\be 
\langle\frac{dE}{dud\Omega}\rangle=-\frac{2G^4M^5}{5R^5c^5}=-\frac{2}{5}\frac{c^5x^5}{G},
\ee where we have defined a dimensionless parameter 
\be 
x=\left(\frac{GM\omega}{c^3}\right)^{2/3}=\frac{GM}{Rc^2}.\label{xR}
\ee In the Newtonian limit, the total energy is 
\be 
E=2\times \frac{1}{2}Mv^2-\frac{GM^2}{2R}=-\frac{1}{4}Mc^2 x.
\ee In the adiabatic approximation, it is sufficient to obtain the variation of $x$ 
\be 
\dot x=\frac{8}{5}\frac{c^3x^5}{GM}
\ee from the energy flux-balance equation. The sign on the right hand side is positive since the velocity of the stars increases  as they revolve around each other. Combining with \eqref{xR}, the result matches the decreasing rate of the radius  in  \cite{Blanchet_2023} at the leading order. 
Note that in the adiabatic approximation, the angular distribution of the helicity flux density doesn't change at the leading order while its magnitude will change. After  substituting the previous result into the expression \eqref{Ecx}, we find the rate of change of the magnitude of helicity flux density  
\bea 
\dot{E}_c=\frac{7}{5\pi}\frac{c^5}{G}x^{15/2}.
\eea Note that the right hand side is positive since the absolute value of the helicity flux density is proportional to $v^7$ and the velocity increases in the process. 

The previous discussion is in the inspiral state where the PN expansion is valid. At the second stage of black hole merger, two black holes form a distorted black hole and the PN expansion becomes invalid. There should be various nonlinear dynamics and one may use numerical relativity to simulate the process. It is rather interesting to understand the helicity flux density at this stage. 

\section{Discussion}\label{dis}
In this work, we have derived the quadrupole formula \eqref{local} for helicity flux density in gravitational radiation and  applied it to the two-body systems in the slow motion and weak field limit. In each case, the total helicity flux on the sphere is always zero while its angle-dependence remains non-trivial. For elliptical orbits, we have computed the average helicity flux density \eqref{helicityelliptic}  over a period. For parabolic or hyperbolic orbits, we have also computed the total helicity flux density during the deflection process. 
In summary, the helicity flux density from a two-body system can always be decomposed as the product of $\cos\theta$ and a function $\zeta(\theta,\phi)$ on the sphere in the quadrupole limit
\be
\frac{dH}{dud\Omega}=\frac{G}{8\pi}\cos\theta \ \zeta(\theta,\phi),\label{fac}
\ee where the function $\zeta(\theta,\phi)$ is determined by the quadrupole moments
\bea 
\zeta(\theta,\phi)&=&3(\dddot{M}_{11}\ddot{M}_{22}-\dddot{M}_{22}\ddot{M}_{11})n_1n_2+(\dddot{M}_{22}\ddot{M}_{12}-\dddot{M}_{12}\ddot{M}_{22})(1+n_1^2-2n_2^2)\nn\\&&+(\dddot{M}_{12}\ddot{M}_{11}-\dddot{M}_{11}\ddot{M}_{12})(1-2n_1^2+n_2^2)
\eea with $n_i$ the $i$-th component of the unit normal vector. In Appendix \ref{cos}, we prove this formula and show that the conclusion is still valid for  a general planar system, where the celestial motion is constrained in a two dimensional plane. Moreover, since $\zeta(\theta,\phi)$ is a parity even and quadratic polynomial of the normal vector, the non-vanishing integrated helicity fluxes \eqref{integrated} are the modes with  $\ell=1$ or $\ell=3$.

We extend the formula to \eqref{total} by including the higher multipoles.  There are various extensions which deserve study in the future.
\begin{enumerate}
    \item In the framework of post-Newtonian (PN) expansion, the radiative multipole moments can be expressed as functionals of the source canonical moments \cite{Blanchet:2013haa}. The PN computation of the energy, linear momentum and angular momentum fluxes has been explored to higher PN orders  \cite{Bini:2022enm}. There are various new effects at higher PN orders, including the radiation reaction correction of the orbits \cite{1985AIHPA..43..107D,Damour:1981bh,Damour:1988mr,1993PhLA..174..196S,Memmesheimer:2004cv}, hereditary effects \cite{Blanchet:1987wq,Blanchet:1992br,Blanchet:1993ec} and so on \cite{Blanchet:2006zz}.  Therefore, it would be better to include  these higher-order corrections in \eqref{corrections} to improve the PN expansions of the helicity flux density.

    \item In electromagnetic theory, there is a similar electromagnetic helicity flux operator in the context of Carrollian holography \cite{Liu:2023qtr}. There are already some discussions on the physical consequences of this electromagnetic helicity flux in \cite{oblak2024orientation}. In astrophysics, one can also define  magnetic helicity  \cite{elsasser1956hydromagnetic,woltjer1958theorem} which describes dynamo processes. The magnetic helicity density $\bm a\cdot \bm b$ is evaluated on a constant time slice $\mathcal{H}$  \footnote{We use $\bm a$ to denote the magnetic vector potential  and $\bm b$ the magnetic field.} while the electromagnetic helicity flux density is defined on $\mathcal{I}^{+}$. In Figure \ref{pen}, we have drawn the hypersurface $\mathcal{H}$ and future null infinity in the Penrose diagram.
    By definition, they are different quantities since they are defined in different hypersurfaces of Penrose diagram. Moreover, the magnetic helicity density is  gauge-dependent \footnote{One may extract a gauge invariant quantity by separating the transverse modes and the longitudinal mode \cite{maleknejad2023photon}.}, while the electromagnetic helicity flux density is gauge-invariant up to large gauge transformations. However, the integrated electromagnetic helicity flux could be equivalent to the magnetic helicity for $g=1$ when there is no topological obstacle between the hypersurface $\mathcal{H}$ and future null infinity $\mathcal{I}^+$. For $g\not=1$, the integrated helicity flux encodes more angle-dependent information than the magnetic helicity flux.
    The method presented here can be extended to the Maxwell field and one may expect a similar multipole formula for the electromagnetic helicity flux density. 
    \begin{figure}
        \centering
        \includegraphics[width=3.9in]{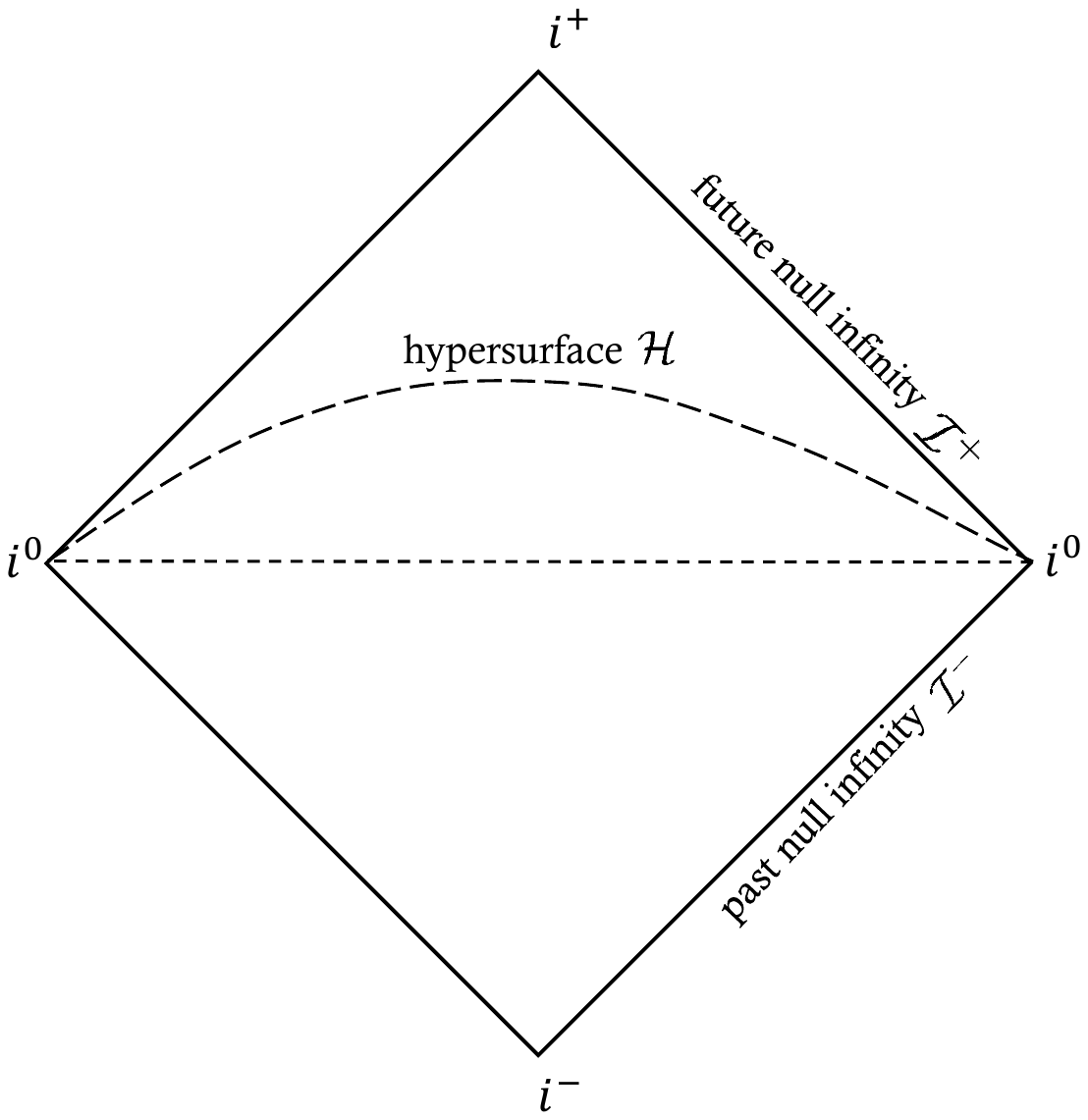}
        \caption{Magnetic helicity density and electromagnetic helicity flux density in Penrose diagram. The former is defined on a constant time slice $\mathcal{H}$ while the latter is defined at future null infinity $\mathcal{I}^+$. The integrated magnetic helicity is equivalent to the integrated electromagnetic helicity flux ($g=1$) when there is no topological obstacle between $\mathcal{H}$ and $\mathcal{I}^+$. }
        \label{pen}
    \end{figure}
    \item In this work, we have discussed the helicity flux density for the orbits in Newtonian mechanics. However, in general relativity, the timelike orbits in the  outer event horizon region in a general Kerr background have been classified \cite{Compere:2021bkk} in the EMRI limit and the orbits are much richer. Therefore, it would be interesting to discuss the helicity flux density for each type of orbits.
    
    \item For real astrophysical systems, the compact stars are extended objects with internal structures that contribute  to the radiative gravitational waves.  Problems such as two coalescing neutron stars \cite{1977ApJ...215..311C}, the black-hole-neutron-star collisions\cite{Lattimer:1974slx}, and the intermediate mass-ratio coalescences \cite{Chen:2019hac} are interesting topics to study since the helicity flux density may also encode the information about the internal structures of neutron stars. 

\end{enumerate}

\vspace{10pt}
{\noindent \bf Acknowledgments.} 
We thank the collaboration of Jinzhuang Dong at the early stage of this work. The work of J.L. was supported by NSFC Grant No. 12005069.

\appendix
\section{Integrals on the unit sphere}\label{integrals}
In this appendix, we will introduce the necessary details on the symmetric trace free Cartesian tensors and their integrals on the unit sphere. The $\ell$-th symmetric trace free Cartesian tensor is defined as
\be 
n^{j(\ell)}=n^{j_1\cdots j_\ell}=n^{j_1}\cdots n^{j_\ell}-\text{traces},
\ee where $n^j$ is the unit normal vector on the unit sphere. More explicitly \cite{Thorne:1980ru}, 
\bea 
n^{j(\ell)}=\sum_{p=0}^{\lfloor\ell/2\rfloor}a(p;\ell)\delta^{(j_{1}j_{2}}\cdots \delta^{j_{2p-1}j_{2p}}n^{j_{2p+1}}\cdots n^{j_{\ell})}\label{cartesian}
\eea with 
\be 
a(p;\ell)=(-1)^{p}\frac{\ell!(2\ell-2p-1)!!}{2^{p}p!(\ell-2p)!(2\ell-1)!!}.
\ee Here the round brackets $(\cdots)$ means that the indices inside the brackets are symmetrized with normalization 1. For example, 
\be 
T^{(ij)}=\frac{1}{2}(T^{ij}+T^{ji}).
\ee For each fixed $\ell$, there are $2\ell+1$ independent symmetric trace free Cartesian tensors $n^{j(\ell)}$ which are related to the $\ell$-th spherical harmonic function $Y_{\ell,m}, m=-\ell,-\ell+1,\cdots,\ell$ by a linear transformation. The properties of the Cartesian tensors $n^{j(\ell)}$ are shown in the following. 
\begin{enumerate}
    \item Parity. The spherical coordinates of the sphere is denoted as 
    \be 
    \Omega=(\theta,\phi).
    \ee Under the inverse transformation that sends the point $\Omega$ to its antipodal point $\Omega^P$,
    \be 
  P: \Omega\to \Omega^P=(\pi-\theta,\pi+\phi),
    \ee  the normal vector $n^j(\Omega)$ flips a sign 
    \be 
    n^j\to -n^j.
    \ee As a consequence, the Cartesian tensor $n^{j(\ell)}$ is parity even for $\ell$ even and parity odd for $\ell$ odd  respectively
    \be 
    P(n^{j(\ell)})=(-1)^\ell n^{j(\ell)}.
    \ee 
   \item Orthogonality. For two Cartesian tensors $n^{i(\ell)}$ and $n^{j(\ell')}$, the integral of their products on the unit sphere is 
   \bea 
   \frac{1}{4\pi}\int d\Omega n^{i(\ell)} n^{j(\ell')}=\frac{\ell!}{(2\ell+1)!!}\Delta^{i(\ell),j(\ell)}\delta_{\ell,\ell'}.
   \eea It vanishes for $\ell\not=\ell'$ and $\Delta^{i(\ell),j(\ell)}$ is the so-called isotropic Cartesian tensor \cite{hess2015tensors}. The isotropic Cartesian tensor $\Delta^{i(\ell),j(\ell)}$ is doubly symmetric traceless in the sense that 
   \bea 
   \Delta^{i_1\cdots i_\ell,j_1\cdots j_\ell}&=&\Delta^{(i_1\cdots i_\ell),j_1\cdots j_\ell}=\Delta^{i_1\cdots i_\ell,(j_1\cdots j_\ell)},\quad\\ \Delta^{i_1\cdots i_\ell,j_1\cdots j_\ell}\delta_{i_1i_2}&=&\Delta^{i_1\cdots i_\ell,j_1\cdots j_\ell}\delta_{j_1j_2}=0.
   \eea The explicit form may be found by combining the fundamental integral on the unit sphere \cite{Thorne:1980ru}
   \bea 
   \frac{1}{4\pi}\int d\Omega n^{j_1}\cdots n^{j_\ell}=\left\{\begin{array}{cc} 0&\ell\ \text{odd},\\
   \frac{1}{\ell+1}\delta_{(j_1j_2}\cdots \delta_{j_{\ell-1}j_\ell)}&\ell \ \text{even}\end{array}\right.\label{integral}
   \eea and the equation \eqref{cartesian}, the result is \cite{Liu:2023jnc}
   \bea 
   \Delta^{i(\ell),j(\ell)}=\sum_{p,q}a(p,q;\ell)\delta^{(i_1i_2}\cdots\delta^{i_{2p-1}i_{2p}}X_{p,q}^{i_{2p+1}\cdots i_\ell),(j_{2q+1}\cdots j_{\ell}}\delta^{j_1j_2}\cdots\delta^{j_{2q-1}j_{2q})},
   \eea where $ X^{i_1\cdots i_\ell,j_1\cdots j_\ell}_{0,0}$ is the doubly symmetric rank $2\ell$ tensor which is constructed by Kronecker signature
    \bea 
   X^{i_1\cdots i_\ell,j_1\cdots j_\ell}_{0,0}= X^{i_1\cdots i_\ell,j_1\cdots j_\ell}=\frac{1}{\ell!}\sum_{\pi\in S_\ell}\delta^{i_1j_{\pi(1)}}\cdots \delta^{i_\ell j_{\pi(\ell)}}
    \eea with $S_\ell$ the group of the permutations of the first $\ell$
natural numbers. Obviously, it is symmetric for the same type of indices
\be 
X^{i_1\cdots i_\ell,j_1\cdots j_\ell}=X^{(i_1\cdots i_\ell),j_1\cdots j_\ell}=X^{i_1\cdots i_\ell,(j_1\cdots j_\ell)}.
\ee The tensor $X_{p,q}^{i_{2p+1}\cdots i_\ell,j_{2q+1}\cdots j_\ell}$ is found by taking the traces $p$ and $q$ times for the $i$ and $j$ indices, respectively
\bea 
X_{p,q}^{i_{2p+1}\cdots i_\ell,j_{2q+1}\cdots j_\ell}=\delta_{i_1i_2}\cdots \delta_{i_{2p-1}i_{2p}}X^{i_1\cdots i_\ell,j_1\cdots j_\ell}_{0,0}\delta_{j_1j_2}\cdots \delta_{j_{2q-1}j_{2q}}.
\eea The coefficient $a(p,q;\ell)$ is the product of $a(p;\ell)$ and $a(q;\ell)$
\be 
a(p,q;\ell)=a(p;\ell)a(q;\ell).
\ee By definition, the isotropic Cartesian tensor $\Delta^{i(\ell),j(\ell)}$ is invariant under the exchange of indices $i(\ell)$ and $j(\ell)$
\be 
\Delta^{i(\ell),j(\ell)}=\Delta^{j(\ell),i(\ell)},
\ee and it may be regarded as a projector which projects any rank $\ell$ tensor to its symmetric trace free part 
\bea 
A^{i(\ell)}=\Delta^{i(\ell),j_1\cdots j_\ell}A^{j_1\cdots j_\ell}.
\eea In particular, 
\be 
n^{i(\ell)}=\Delta^{i(\ell),i_1'\cdots i_\ell'}n^{i_1'}\cdots n^{i_{\ell}'}=\Delta^{i(\ell),i'(\ell)}n^{i'(\ell)}.\label{projn}
\ee 
\item Completeness relation. For two  Cartesian tensors $n^{i(\ell)}$ with different arguments, we have the summation 
\be 
\frac{1}{4\pi}\sum_{\ell=0}^\infty \frac{(2\ell+1)!!}{\ell!}n^{i(\ell)}(\Omega)n^{i(\ell)}(\Omega')=\delta(\Omega-\Omega').\label{complete0}
\ee To prove this relation, we need the completeness relation of spherical harmonic functions 
\be 
\sum_{\ell=0}^\infty \sum_{m=-\ell}^\ell Y_{\ell,m}(\Omega)Y^*_{\ell,m}(\Omega')=\delta(\Omega-\Omega').\label{complete}
\ee With the addition theorem of the spherical harmonic function 
\bea 
P_{\ell}(\cos\gamma)=\frac{4\pi}{2\ell+1}\sum_{m=-\ell}^\ell Y_{\ell,m}(\Omega)Y^*_{\ell,m}(\Omega'),
\eea we may rewrite the relation \eqref{complete} as 
\be 
\frac{1}{4\pi}\sum_{\ell=0}^\infty (2\ell+1)P_{\ell}(\cos\gamma)=\delta(\Omega-\Omega').\label{complete2}
\ee Note that $\gamma$ is the angle between the two normal vectors 
$\bm n=\bm n(\Omega)$ and $\bm n'=\bm n(\Omega')$
\be 
\bm n\cdot\bm n'=\cos\gamma.
\ee The last ingredient is the addition theorem \cite{hess2015tensors} associated with the symmetric trace free tensor $n^{i(\ell)}$
\bea 
P_\ell(\cos\gamma)=\frac{(2\ell-1)!!}{\ell!}n^{i(\ell)}(\Omega)n^{i(\ell)}(\Omega').\label{complete3}
\eea Substituting \eqref{complete3} into \eqref{complete2}, we find the completeness relation \eqref{complete0} for the symmetric trace free tensors $n^{i(\ell)}$. 
With the completeness relation, we may expand functions on the unit sphere as 
\bea 
f(\Omega)=f_{i(\ell)}n^{i(\ell)}(\Omega)
\eea where $f_{i(\ell)}$ is symmetric trace free 
\be 
f_{i(\ell)}=\frac{1}{4\pi }\frac{(2\ell+1)!!}{\ell!}\int d\Omega f(\Omega)n^{i(\ell)}(\Omega).
\ee For spherical harmonic function, 
\bea 
Y_{\ell,m}(\Omega)=\mathscr{Y}^{\ell,m}_{j(\ell)}n^{j(\ell)},
\eea we have 
\bea 
\mathscr{Y}^{\ell,m}_{j(\ell)}=\frac{1}{4\pi}\frac{(2\ell+1)!!}{\ell!}\int d\Omega Y_{\ell,m}(\Omega)n^{j(\ell)}(\Omega).
\eea 
\item Clebsch-Gordan tensors. Similar to the definition of Clebsch-Gordan coefficients, the Clebsch-Gordan tensors are defined by the integral of three symmetric trace free tensors $n^{i(\ell)}$ 
\bea 
\Delta^{i(\ell_1),j(\ell_2),k(\ell_3)}=\frac{1}{4\pi}\int d\Omega\  n^{i(\ell_1)}n^{j(\ell_2)}n^{k(\ell_3)}.
\eea After some efforts, we find 
\bea 
\Delta^{i(\ell_1),j(\ell_2),k(\ell_3)}&=&m_{h_1,h_2,h_3}\Delta^{i(h_2)\bar{i}(h_3),i'(h_2)\bar{i}'(h_3)}\Delta^{j(h_1)\bar{j}(h_3),j'(h_1)\bar{i}'(h_3)}\Delta^{k(h_1)\bar{k}(h_2),j'(h_1)i'(h_2)}\nn\\&&\times\Theta_{h_1}\Theta_{h_2}\Theta_{h_3}.\label{deltaell123}
\eea Here the symbol $\Theta_h$ is similar to the step function. It equals to 1 for non-negative integers and 0 otherwise
\bea 
\Theta_h=\left\{\begin{array}{cc}1&h=0,1,2,\cdots,\\
0&\text{others}.\end{array}\right.
\eea The value of the coefficient $m_{h_1,h_2,h_3}$ is
\bea 
m_{h_1,h_2,h_3}=\frac{\ell_1!\ell_2!\ell_3!}{h_1!h_2!h_3!(\ell_1+\ell_2+\ell_3+1)!!}.\eea To prove the formula \eqref{deltaell123}, we may use the identity \eqref{projn} and the integral \eqref{integral}
\bea 
\Delta^{i(\ell_1),j(\ell_2),k(\ell_3)}&=&\frac{1}{4\pi}\Delta^{i(\ell_1),i'(\ell_1)}\Delta^{j(\ell_2),j'(\ell_2)}\Delta^{k(\ell_3),k'(\ell_3)}\int d\Omega n^{i'_1}\cdots n^{i'_{\ell_1}}n^{j'_1}\cdots n^{j'_{\ell_2}}n^{k'_1}\cdots n^{k'_{\ell_3}}\nn\\&=&\frac{1}{(\ell_1+\ell_2+\ell_3+1)!!}\Delta^{i(\ell_1),i'(\ell_1)}\Delta^{j(\ell_2),j'(\ell_2)}\Delta^{k(\ell_3),k'(\ell_3)}\nn\\&&\times\delta_{(i_1'i_2'}\cdots \delta_{k'_{\ell_3-1} k'_{\ell_3})}.
\eea In the second step, we have assumed the summation $\ell_1+\ell_2+\ell_3$ is even. Since $\Delta^{i(\ell_1),j(\ell_2),k(\ell_3)}$ is triplely symmetric traceless 
\bea
&&\Delta^{i(\ell_1),j(\ell_2),k(\ell_3)}=\Delta^{i_1\cdots i_{\ell_1},j(\ell_2),k(\ell_3)}=\Delta^{i(\ell_1),j_1\cdots j_{\ell_2},k(\ell_3)}=\Delta^{i(\ell_1),j(\ell_2),k_1\cdots k_{\ell_3}},\\
&& \Delta^{i_1i_2\cdots i_{\ell_1},j(\ell_2),k(\ell_3)}\delta_{i_1i_2}=\Delta^{i(\ell_1),j_1\cdots j_{\ell_2},k(\ell_3)}\delta_{j_1j_2}=\Delta^{i(\ell_1),j(\ell_2),k_1\cdots k_{\ell_3}}\delta_{k_1k_2}=0,
\eea the non-trivial contributions are from the contractions among $(i'j'), (j'k')$ or $(k'i')$ indices. In other words, we may split the indices as 
\be 
i(\ell_1)=i(h_2)\bar{i}(h_3),\quad j(\ell_2)=j(h_1)\bar{j}(h_3),\quad k(\ell_3)=k(h_1)\bar{k}(h_2).
\ee  Then the number of contractions between $i'$ and $j'$ indices is $h_3$, and so on. It follows that 
\bea 
\ell_1=h_2+h_3,\quad \ell_2=h_1+h_3,\quad \ell_3=h_1+h_2.
\eea The constants $h_1,h_2,h_3$ are fixed to 
\bea 
h_1=\frac{\ell_2+\ell_3-\ell_1}{2},\quad h_2=\frac{\ell_1+\ell_3-\ell_2}{2},\quad h_3=\frac{\ell_1+\ell_2-\ell_3}{2}.
\eea Therefore, 
\bea 
&&\Delta^{i(\ell_1),j(\ell_2),k(\ell_3)}\nn\\&=&\frac{1}{(\ell_1+\ell_2+\ell_3+1)!!}\Delta^{i(h_2)\bar{i}(h_3),i'(h_2)\bar{i}'(h_3)}\Delta^{j(h_1)\bar{j}(h_3),j'(h_1)\bar{j}'(h_3)}\nn\Delta^{k(h_1)\bar{k}(h_2),k'(h_1)\bar{k}(h_2)}\nn\\&&\times C_{\ell_1}^{h_1}C_{\ell_2}^{h_2}C_{\ell_3}^{h_3}h_1!h_2!h_3!\delta^{i_1'\bar{k}_1'}\cdots \delta^{i_{h_2}'\bar{k}_{h_2}'}\delta^{\bar{i}'_1\bar{j}_1'}\cdots \delta^{\bar{i}'_{h_3}\bar{j}'_{h_3}}\delta^{j_1'k_1'}\cdots\delta^{j'_{h_1}k'_{h_1}}\Theta_{h_1}\Theta_{h_2}\Theta_{h_3}\nn\\&=&m_{h_1,h_2,h_3}\Delta^{i(h_2)\bar{i}(h_3),i'(h_2)\bar{i}'(h_3)}\Delta^{j(h_1)\bar{j}(h_3),j'(h_1)\bar{i}'(h_3)}\Delta^{k(h_1)\bar{k}(h_2),j'(h_1)i'(h_2)}\Theta_{h_1}\Theta_{h_2}\Theta_{h_3}.\nn\\
\eea In the first step, the factor $C_{\ell_1}^{h_1}C_{\ell_2}^{h_2}C_{\ell_3}^{h_3}h_1!h_2!h_3!$ is the number of terms which contribute to the contractions.
\end{enumerate}
In the following, we will use the previous properties to compute several integrals which are relevant to work. We expand the functions $f(\Omega),\ g(\Omega),\ h(\Omega)$ with the symmetric trace free Cartesian tensors
 \bea 
    f(\Omega)=f_{i(\ell)}n^{i(\ell)},\quad g(\Omega)=g_{i(\ell)}n^{i(\ell)},\quad h(\Omega)=h_{i(\ell)}n^{i(\ell)},
    \eea 
   and the corresponding integral properties on the unit sphere as follows
\begin{enumerate}
    \item The product of $f(\Omega)$ and $g(\Omega)$
    \bea 
   \frac{1}{4\pi}\int d\Omega f(\Omega)g(\Omega)=\frac{1}{4\pi}\int d\Omega f_{i(\ell)}n^{i(\ell)} g_{j(\ell')}n^{j(\ell')}=\frac{\ell!}{(2\ell+1)!!}f_{i(\ell)}g_{i(\ell)}.
    \eea 
    \bea 
   && \frac{1}{4\pi}\int d\Omega f(\Omega)g(\Omega)h(\Omega)=\Delta^{i(\ell_1),j(\ell_2),k(\ell_3)}f_{i(\ell_1)}g_{j(\ell_2)}h_{k(\ell_3)}\nn\\&=&m_{h_1,h_2,h_3}f_{i(h_2)\bar{i}(h_3)}g_{j(h_1)\bar{j}(h_3)}h_{k(h_1)\bar{k}(h_2)}\Theta_{h_1}\Theta_{h_2}\Theta_{h_3}.
    \eea The result is consistent with the one in \cite{Compere:2019gft}.  See also similar discussions in \cite{Faye:2014fra}.
\end{enumerate}
In the total helicity flux, the integrand separates into four parts, $\dot U U, \dot U V, \dot V U$ and $\dot V V$. We will discuss them term by term. 
\begin{enumerate}
    \item $\dot U U $ terms. Their contributions to the total helicity flux are always zero. We will take the integral 
    \bea 
    I_1=\frac{1}{4\pi}\int d\Omega \epsilon^{ij'm}\delta^{ji'}n_m n^{i(\ell-2)}n^{i'(\ell'-2)}\dot{U}_{ij i(\ell-2)}U_{i'j'i'(\ell'-2)}
    \eea as an example. Note that the integral is actually 
    \bea 
    I_1=\frac{1}{4\pi}\int d\Omega \epsilon^{ij'm}n_m n^{i(\ell-2)}n^{i'(\ell'-2)}\dot{U}_{iji(\ell-2)}U_{jj'i'(\ell'-2)}.
    \eea After the integrating on the sphere, the index $m$ is either equal to $i'_k,\ k=1,2,\cdots, \ell-2$ or equal to $j_k,\ k=1,2,\cdots,\ell'-2$. However, since the Levi-Civita tensor is antisymmetric and $U_{i(\ell)}$ is symmetric trace free, the result is always zero.
    \item $\dot V V$ terms. Their contributions to the total helicity flux are also zero. We will take the integral 
    \bea 
    I_2=\frac{1}{4\pi}\int d\Omega \epsilon^{ij'm}n^j n^{i'}n_m n^{i(\ell-2)}n^{i'(\ell'-2)}\epsilon_{ipq}n_p\dot{V}_{jq i(\ell-2)}\epsilon_{i'p'q'}n_{p'}V_{j'q'i'(\ell'-2)}
    \eea as an example.  The integral is zero due to the antisymmetric property of Levi-Civita tensor  
    \be 
    \epsilon_{i'p'q'}n_{i'}n_{p'}=0.
    \ee
    \item $\dot U V$ and $\dot V U$ terms. We may choose the integral 
    \bea 
    I_3=\frac{1}{4\pi}\int d\Omega \epsilon^{ij'm}\delta^{ji'}n_m n^{i(\ell-2)}n^{i'(\ell'-2)}\dot{U}_{ij i(\ell-2)}\epsilon_{i'p'q'}n_{p'}V_{j'q'i'(\ell'-2)}
    \eea as an example. 
    Due to the identity
    \be 
    \epsilon_{ij'm}\epsilon_{i'p'q'}=\delta_{ii'}\delta_{j'p'}\delta_{mq'}+\delta_{ip'}\delta_{j'q'}\delta_{mi'}+\delta_{iq'}\delta_{j'i'}\delta_{mp'}-\delta_{ii'}\delta_{j'q'}\delta_{mp'}-\delta_{ip'}\delta_{j'i'}\delta_{mq'}-\delta_{iq'}\delta_{j'p'}\delta_{mi'},
    \ee we may simplify the integral to 
    \bea 
    I_3&=&\frac{1}{4\pi}\int d\Omega n_m n_{p'}n^{i(\ell-2)}n^{i'(\ell'-2)}\dot{U}_{ii'i(\ell-2)}V_{j'q'i'(\ell'-2)}\nn\\&\times&(\delta_{iq'}\delta_{j'i'}\delta_{mp'}-\delta_{ip'}\delta_{j'i'}\delta_{mq'}-\delta_{iq'}\delta_{j'p'}\delta_{mi'})\nn\\&=&(m_{\ell-2}-2m_{\ell-1})\dot{U}_{i(\ell)}V_{i(\ell)},
    \eea where 
    \bea
    m_\ell=\frac{\ell!}{(2\ell+1)!!}.
    \eea
\end{enumerate}

\section{Quadrupole formula}\label{cos}
In this section, we will prove \eqref{fac} for planar systems. For a general $N$-body system, the quadrupole moment is still parameterized by a symmetric trace free matrix $M_{ij}$
\bea 
M_{ij}=\left(\begin{array}{ccc}M_{11}&M_{12}&M_{13}\\ M_{12}&M_{22}&M_{23}\\ M_{13}&M_{23}&M_{33}\end{array}\right),
\eea where the component $M_{33}$ is not independent 
\be 
M_{33}=-M_{11}-M_{22}.
\ee Now we can expand the qudrupole formula \eqref{local} straightforwardly 
\bea 
\frac{dH}{du d\Omega}&=&\frac{G}{8\pi}[\dddot{M}_{13}(\ddot{M}_{11} \left(n_1^2-1\right) {n_2}+\ddot{M}_{12} {n_1} \left(-n_1^2+n_2^2+n_3^2+1\right)-\ddot{M}_{22} n_1^2 {n_2}\nn\\&&+\ddot{M}_{22} {n_2} n_3^2-\ddot{M}_{23} n_1^2 {n_3}-\ddot{M}_{23} n_2^2 {n_3}+\ddot{M}_{23} n_3^3-\ddot{M}_{23} {n_3}-\ddot{M}_{33} {n_2} n_3^2+\ddot{M}_{33} {n_2})\nn\\&&+\dddot{M}_{23}(\ddot{M}_{11} {n_1} n_2^2-\ddot{M}_{11} {n_1} n_3^2-\ddot{M}_{12} n_1^2 {n_2}+\ddot{M}_{12} {n_2} \left(n_2^2-1\right)-\ddot{M}_{12} {n_2} n_3^2\nn\\&&+\ddot{M}_{13} n_1^2 {n_3}+\ddot{M}_{13} n_2^2 {n_3}+\ddot{M}_{13} \left({n_3}-n_3^3\right)+\ddot{M}_{22} \left({n_1}-{n_1} n_2^2\right)+\ddot{M}_{33} {n_1} \left(n_3^2-1\right))\nn\\&&+\ddot{M}_{13}(\dddot{M}_{11} \left({n_2}-n_1^2 {n_2}\right)+\dddot{M}_{12} {n_1} \left(n_1^2-n_2^2-n_3^2-1\right)+\dddot{M}_{22} n_1^2 {n_2}-\dddot{M}_{22} {n_2} n_3^2\nn\\&&+\dddot{M}_{23} n_1^2 {n_3}+\dddot{M}_{23} n_2^2 {n_3}-\dddot{M}_{23} n_3^3+\dddot{M}_{23} {n_3}+\dddot{M}_{33} {n_2} n_3^2-\dddot{M}_{33} {n_2})\nn\\&&+\ddot{M}_{23}(-\dddot{M}_{11} {n_1} n_2^2+\dddot{M}_{11} {n_1} n_3^2+\dddot{M}_{12} n_1^2 {n_2}+\dddot{M}_{12} \left({n_2}-n_2^3\right)+\dddot{M}_{12} {n_2} n_3^2\nn\\&&-\dddot{M}_{13} n_1^2 {n_3}-\dddot{M}_{13}n_2^2 {n_3}+\dddot{M}_{13} {n_3} \left(n_3^2-1\right)+\dddot{M}_{22} {n_1} \left(n_2^2-1\right)+\dddot{M}_{33} \left(n_1-{n_1} n_3^2\right))\nn\\&&-(\ddot{M}_{13}\dddot{M}_{23}-\ddot{M}_{23}\dddot{M}_{13})({n_3} \left(n_1^2+n_2^2-n_3^2+1\right))]\nn\\&&+\frac{G}{8\pi}n_3[3(\dddot{M}_{11}\ddot{M}_{22}-\dddot{M}_{22}\ddot{M}_{11})n_1n_2+(\dddot{M}_{22}\ddot{M}_{12}-\dddot{M}_{12}\ddot{M}_{22})(1+n_1^2-2n_2^2)\nn\\&&+(\dddot{M}_{12}\ddot{M}_{11}-\dddot{M}_{11}\ddot{M}_{12})(1-2n_1^2+n_2^2)].
\eea 
Now that the right hand side is not proportional to $n_3$ in general. However, for planar systems, one can always construct a coordinate system such that  $z=0$ corresponds to the plane. From the definition of the quadupole moments, we have 
\bea 
M_{13}=M_{23}=0
\eea for the planar systems. Then all the terms that contain $M_{13}$ or $M_{23}$ are zero. The quadrupole formula becomes 
\bea 
\frac{dH}{du d\Omega}&=&\frac{G}{8\pi}n_3[3(\dddot{M}_{11}\ddot{M}_{22}-\dddot{M}_{22}\ddot{M}_{11})n_1n_2+(\dddot{M}_{22}\ddot{M}_{12}-\dddot{M}_{12}\ddot{M}_{22})(1+n_1^2-2n_2^2)\nn\\&&+(\dddot{M}_{12}\ddot{M}_{11}-\dddot{M}_{11}\ddot{M}_{12})(1-2n_1^2+n_2^2)],
\eea which is exactly \eqref{fac}. Interestingly, one can also obtain the integrated helicity flux for periodic oribts using \eqref{inthe}
\bs\begin{align}
   \langle \mathcal{O}_{1,0}\rangle&=\frac{G}{5 \sqrt{3 \pi }}\langle -\dddot{M}_{11} \ddot{M}_{12}+\ddot{M}_{11} \dddot{M}_{12}-\dddot{M}_{12} \ddot{M}_{22}+\ddot{M}_{12} \dddot{M}_{22}\rangle,\\
   \langle\mathcal{ O}_{3,0}\rangle&=\frac{G}{20 \sqrt{7 \pi }}\langle -\dddot{M}_{11} \ddot{M}_{12}+\ddot{M}_{11} \dddot{M}_{12}-\dddot{M}_{12} \ddot{M}_{22}+\ddot{M}_{12} \dddot{M}_{22}\rangle,\\
   \langle\mathcal{ O}_{3,2}\rangle&=\frac{G}{4} \sqrt{\frac{3}{70 \pi }} \langle\dddot{M}_{11} (\ddot{M}_{12}+i \ddot{M}_{22})-\ddot{M}_{11} (\dddot{M}_{12}+i \dddot{M}_{22})-\dddot{M}_{12} \ddot{M}_{22}+\ddot{M}_{12} \dddot{M}_{22}\rangle,\\
    \langle\mathcal{O}_{3,-2}\rangle&=\langle\mathcal{O}_{3,2}^*\rangle.
\end{align}\es 

\bibliography{refs}
\end{document}